\documentclass[12pt]{article}
\pdfoutput=1
\usepackage{jheppub}
\usepackage{breakurl}
\usepackage{mathrsfs}
\usepackage{amsmath}
\usepackage{amsfonts}
\usepackage{mathrsfs}
\usepackage{slashed}
\usepackage{epigraph}
\usepackage{fancyhdr}
\usepackage{amssymb}
\usepackage{graphicx}
\usepackage{longtable}
\usepackage{makecell}
\usepackage{amscd}
\usepackage{tabu}
\usepackage{bm}
\usepackage{multirow}
\usepackage{xspace}
\usepackage{color}
\usepackage{hyperref} 
\hypersetup{linktocpage=true}
\usepackage[all]{hypcap}
\hypersetup{
   bookmarks=true,         
   unicode=false,          
   pdftoolbar=true,        
   pdfmenubar=true,        
   pdffitwindow=false,     
   pdfstartview={FitH},    
   pdftitle={My title},    
   pdfauthor={Author},     
   pdfsubject={Subject},   
   pdfcreator={Creator},   
   pdfproducer={Producer}, 
   pdfkeywords={keyword1} {key2} {key3}, 
   pdfnewwindow=true,      
   colorlinks=true,        
   linkcolor=blue,         
   citecolor=blue,      
   filecolor=magenta,      
   urlcolor=blue          
}
\usepackage{fancyhdr}
\def\beq{\begin{equation}} 
\def\eeq{\end{equation}} 
\def\bea{\begin{eqnarray}} 
\def\eea{\end{eqnarray}} 
 
\def\tr{\textcolor{red}}

\def\bc {\begin{center}}
\def\ec {\end{center}}           
\def\stopl{\widetilde{t}_1}
\def\stoph{\widetilde{t}_2}
\def\sbottom1{\widetilde{b}_1}
\def\charginol{\widetilde{\chi}_1^\pm}
\def\neutl{\widetilde{\chi}_1^0}
\def\neuth{\widetilde{\chi}_2^0}

\def\TeV{\, {\rm TeV}}
\def\GeV{\, {\rm GeV}}

\def\mstl{m_{\widetilde{t}_1}}
\def\msth{m_{\widetilde{t}_2}}
\def\msbl{m_{\widetilde{b}_1}}
\def\mlsp{m_{\widetilde{\chi}_1^0}}
\def\mc1{m_{\widetilde{\chi}_1^\pm}}
\def\Zhad{Z^*_{had}}
\def\Whad{W^*_{had}}
\def\Wlep{W^*_{lep}}
\def\Zlep{Z^*_{lep}}
\newcommand{\br}{\mbox{\ensuremath{\mathcal{B}}}}
\newcommand{\sigmaXBF}{\mbox{\ensuremath{\sigma\times\mathcal{B}}}\xspace}
\def\smodels  {{\sc SModelS}\,}

\newcommand{\comment}[1]{}
\def\lsim{\buildrel{\scriptscriptstyle <}\over{\scriptscriptstyle\sim}}

\title{Light stop in the MSSM after LHC Run 1}
\author[a,1]{Genevi\`eve B\'elanger}
\author[b,c,2]{Diptimoy Ghosh}
\author[d,3]{Rohini Godbole}
\author[e,f,4]{Suchita Kulkarni}
\affiliation[a]{\normalfont{LAPTH, Universit\'e Savoie Mont Blanc, CNRS, B.P.110, 
F-74941 Annecy-le-Vieux, France}}
\affiliation[b]{\normalfont{INFN, Sezione di Roma, Piazzale A. Moro 2, I-00185 Roma, Italy}}
\affiliation[c]{\normalfont{Department of Particle Physics and Astrophysics, 
Weizmann Institute of Science, Rehovot 76100, Israel}}
\affiliation[d]{\normalfont{Center for High Energy Physics, Indian Institute of Science,
Bangalore, 560012, India}}
\affiliation[e]{\normalfont{Laboratoire de Physique Subatomique et de Cosmologie, 
Universit\'e Grenoble-Alpes, CNRS/IN2P3, 53 avenue des Martyrs, F-38026 Grenoble Cedex, 
France}}
\affiliation[f]{\normalfont{Institut f\"ur Hochenergiephysik,  
\"Osterreichische Akademie der Wissenschaften, Nikolsdorfer Gasse 18, 1050 Wien, Austria}}
\emailAdd{genevieve.belanger@lapth.cnrs.fr}
\emailAdd{diptimoy.ghosh@weizmann.ac.il}
\emailAdd{rohini@cts.iisc.ernet.in} 
\emailAdd{suchita.kulkarni@oeaw.ac.at}

\abstract{The discovery of a Higgs boson with a mass of 126 GeV at the LHC when combined with the 
non-observation of new physics both in direct and indirect searches imposes strong constraints on 
supersymmetric models and in particular on the top squark sector. The experiments for direct 
detection of dark matter have provided with yet more constraints on the neutralino LSP mass and its 
interactions. After imposing limits from the Higgs, flavour and dark matter sectors, we examine 
the feasibility for a light stop in the context of the pMSSM, in light of current results for stop 
and other SUSY searches at the LHC. We only require that the neutralino dark matter explains a 
fraction of the cosmologically measured dark matter abundance. We find that a stop with mass below $\sim$ 
500 GeV is still allowed. We further study various probes of the light stop scenario that could be performed at the 
LHC Run - II either through direct searches for the light and heavy stop, or SUSY searches not currently 
available in simplified model results. Moreover we study the characteristics of heavy Higgs for the points in 
the parameter space allowed by all the available constraints and  
illustrate the region  with large cross sections to fermionic or electroweakino channels. 
Finally we show that nearly all scenarios with a small stop$-$LSP mass 
difference will be tested by Xenon1T provided the NLSP is a chargino, thus probing a region hard 
to access at the LHC.}

\begin{document} 
\begin{flushright}
LAPTH-030/15 \\
LPSC15132\\
HEPHY-PUB 953/15
\end{flushright}
\maketitle
\flushbottom
\section{Introduction}

The Large Hadron Collider (LHC) has completed its first run (Run-I) with an unprecedented success. 
A Higgs particle has been discovered with a mass $\sim$ 126 GeV~\cite{Chatrchyan:2012ufa, 
Aad:2012tfa,Chatrchyan:2013lba,Aad:2013xqa,Aad:2015zhl,Khachatryan:2014jba}. Its couplings to the 
Standard Model (SM) electroweak gauge bosons have been established to be close to the SM 
expectations by measurements of signal rates~\cite{Khachatryan:2014jba,Chatrchyan:2013iaa,
Chatrchyan:2013zna,Aad:2014lma,Aad:2014eva, 
Aad:2015gra,Khachatryan:2014qaa,Aad:2014xzb,Chatrchyan:2014nva,ATLAS-CONF-2014-060} and 
spin-parity determinations~\cite{Aad:2015rwa,Chatrchyan:2013mxa,Chatrchyan:2012jja,Chatrchyan:2013iaa}. 
While there remains 
considerable room for deviations in the couplings to fermions, no sign of any New Physics (NP) has 
been detected yet. The search for new states near the electroweak scale has also been frustrating. 
The trudge of null results has shrunk the parameter spaces of weak scale NP models considerably, 
and models of weak scale Supersymmetry (SUSY) are not exceptions.

Indeed, the general tone at the moment is fairly lugubrious for SUSY enthusiasts. The simplest 
versions of the Minimal Supersymmetric Standard Model (MSSM), with the simplest assumptions about 
the high-scale theory, are under increasing tension with a wide range of experimental data 
including the $\sim$ 126 GeV mass of the Higgs boson~\cite{Arbey:2011ab,Ghosh:2012dh,Dighe:2013wfa,
Buchmueller:2013rsa}. However, the above argument 
can be turned around to advocate many convincing reasons to study low energy SUSY. For example, 
that the electroweak symmetry is broken by an elementary scalar whole mass is below 135 GeV, is, 
in fact, a prediction of the MSSM. One should also be reminded that the MSSM has 
excellent decoupling properties which keep electroweak precision observables under control and to 
some extent also ameliorate the tension with limits on flavour observables. Moreover, the states 
which are directly related to the naturalness of Electro-Weak Symmetry Breaking (EWSB) are not 
constrained severely by the direct searches yet.

Existence of these states around the TeV scale is implied by demands of naturalness, which is a 
much discussed issue in the context of Supersymmetric theories (for example see~\cite{Martin:1997ns,
Drees:2004jm}). The main point can be understood by considering the issue of
stabilization of the Higgs mass against radiative corrections. To be specific, the correction to the 
Higgs mass due to radiative effects can be 
written as,
\bea
\delta m_h^2 (\Lambda_{\rm EW}) \sim \Lambda_{\rm SUSY}^2 \, \ln\left( \dfrac{\Lambda_{\rm mess}}
{\Lambda_{\rm EW}} \right) \, \, ,
\label{eq-2}
\eea
where $\Lambda_{\rm mess}$ denotes the scale at which SUSY breaking effects are mediated to the 
MSSM and a common mass scale $\Lambda_{\rm SUSY}$ for all the SUSY particles has been assumed. 
Eq.~(\ref{eq-2}) immediately makes it clear why SUSY particles (especially those which couple 
strongly to the Higgs) $\lesssim$ TeV are desired.  

For moderate to large $\tan\beta \equiv \langle H_u \rangle/\langle H_d \rangle$, e.g. 
$\tan\beta \gtrsim 2$, the Higgs mass in the MSSM can be written as~\cite{Martin:1997ns}
\beq
m_h^2 = -2 (|\mu|^2 + m_{H_u}^2|_{\rm tree} + m_{H_u}^2|_{\rm rad}),
\label{eq-3}
\eeq
where $\mu$ is the supersymmetric Higgs mass parameter, and $m_{H_u}^2|_{\rm tree}$ and 
$m_{H_u}^2|_{\rm rad}$ are the tree-level and radiative contributions to the soft SUSY breaking 
mass squared for $H_u$. The dominant radiative correction to $m_{H_u}^2$ proportional to the 
top quark Yukawa coupling is given by~\cite{Martin:1997ns},
\beq
m_{H_u}^2|_{\rm rad} \simeq -\frac{3y_t^2}{8\pi^2} \bigl( m_{\widetilde{Q}_3}^2 + 
m_{\widetilde{U}_3}^2 + |A_t|^2 \bigr) \ln\Biggl( \frac{\Lambda_{\rm mess}}{\rm M_{\widetilde{t}}} 
\Biggr),
\label{eq-4}
\eeq
where $y_t$ is the top Yukawa coupling, $m_{\widetilde{Q}_3}^2$ and $m_{\widetilde{U}_3}^2$ are the 
soft SUSY breaking mass squared parameters for the third-generation squark doublet and singlet up-type 
squark, $A_t$ is the scalar  trilinear interaction parameter for the top squarks
\footnote{Note that we use stop and top squark interchangeably.} and 
$\rm M_{\widetilde{t}}$ denotes an average mass scale for the top squarks.  
Recall that in Supersummetric theories a light Higgs is `natural' in the sense that the stabilization 
of the Higgs mass around the EW scale is guaranteed by the symmetry. The destabilizing effects come 
from SUSY breaking. If none of the terms on the right-hand-side of Eq.~\ref{eq-3} are much larger than 
the left-hand-side then it implies that no fine tuning of parameters in the theory is needed to guarantee 
the low Higgs mass. Thus the amount of `cancellations' (fine tuning) required to satisfy Eq.~\ref{eq-3} 
is then a measure of `naturalness'. As an  example if we define the fine-tuning parameter 
$\Delta = 2m_{H_u}^2|_{\rm rad}/m_h^2$ one gets an upper limit on the top squark mass scale as a function 
of the fine-tuning parameter~\cite{Ellis:1986yg,Barbieri:1987fn,Kitano:2006gv},
\bea
\widetilde{A_t} \equiv \sqrt{\mstl^2 + \msth^2 + A_t^2}  & 
\lesssim & 600\GeV \, 
\sqrt{\dfrac{3}{\ln(\Lambda_{\rm mess}/\TeV)}} \, \sqrt{\dfrac{\Delta}{5}} \, \, .
\label{stop-naturalness}
\eea
Upper bounds on $\mu$ and the gluino mass can also be obtained in a similar way,
\bea
\mu  & \lesssim & 200 \GeV \, \sqrt{\dfrac{\Delta}{5}} \, \, ,\\
M_{3} &\lesssim& 900\GeV \,\left(\dfrac{3}{\ln(\Lambda_{\rm mess}/\TeV)} \right)
\, \sqrt{\dfrac{\Delta}{5}} \, \, \label{gluino-tuning}.
\eea
Equation~\ref{gluino-tuning} 
follows from taking into account, in the leading-logarithm (LL) approximation, corrections to $m_h^2$ coming from 
gluino mass. These come from the gluino induced corrections to the Higgs potential arising at two loops which in 
turn come from gluino corrections to the stop mass at one loop.

If we forget about the Higgs mass and other indirect constraints for a moment, then from direct 
searches alone one can still have $\mstl \sim \msth \sim 500$ GeV 
and $A_t \sim 0$. This amounts to a tuning $\Delta < 10$ (assuming $\Lambda_{\rm mess} = 20$ 
TeV). However, the direct search bound on the gluino mass $m_{\widetilde{g}} \gtrsim 1500$ GeV 
requires, according to Eq.~\ref{gluino-tuning}, the tuning to be $\Delta \gtrsim 15$. This means 
that, as far as the direct search bounds are concerned, it is the gluino mass that has stronger 
effect on fine tuning than the stop mass. Of course, when constraints from the Higgs mass are taken 
into account, low values for $A_t$ are not allowed (if at least one of the top squarks is desired to 
be light). This makes the tuning much worse ($\Delta > 50$).

The effect of gluino mass on fine-tuning can also be understood if one considers the running of the
stop mass. Indeed the top squark is a scalar and its mass is subject to the same fine tuning problem as 
the Higgs mass. More precisely, the leading contribution is given by~\cite{Martin:1997ns}
\bea
\dfrac{d m_{\widetilde{t}}^2}{d \ln(\mu)} = - \dfrac{1}{16 \pi^2} \dfrac{32}{3} g_3^2 M_3^2 \, \, .
\label{eq-5}
\eea
This means that the stop mass is attracted towards the gluino mass at low energies and thus a 
light stop does not seem very `natural' in view of the rather high lower bounds on the gluino mass implied by LHC Run-I data. One way to allow a light stop and still be consistent with 
naturalness, is to somehow weaken the rather strong bounds on the gluino mass implied by the 
current LHC data.  This can be done in SUSY models like compressed SUSY, stealth SUSY and 
R-parity violating SUSY, hence the  renewed interest in such models~\cite{LeCompte:2011cn,
Fan:2011yu,Han:2012cu,Evans:2012bf,Berger:2013sir,Biswas:2013hfa}.

One must further appreciate that the naturalness criteria, to some extent, tend to be subjective. 
In fact, Ref.~\cite{Baer:2012mv}, argues that an appropriate model-independent measure of fine 
tuning is not in terms of the high scale quantities, but rather in terms of $\mu$. Thus a light 
Higgsino is the one robust demand one can make on the particle spectrum by requiring naturalness, 
without making any assumptions about the high scale physics.  In this analysis they construct a 
measure of naturalness which is `independent' of the precise model of the high scale physics. This 
measure can be small even when the conventional measures of fine tuning such as $\Delta$ take 
large values. Due to such subjectivity in the naturalness criteria, in our analysis we will cover 
values of parameters beyond the nominal upper bounds indicated by 
Eqs.~\ref{stop-naturalness}--\ref{gluino-tuning}, for a given value of $\Delta$.

Finally, irrespective of the amount of fine tuning introduced it is interesting to investigate 
to which extent light stop masses are allowed by all the existing direct and indirect 
measurements, and to see how well this region can be tested at the LHC. Here, we will work within 
the framework of the CP and flavour conserving phenomenological MSSM (pMSSM) with 19 free 
parameters~\cite{Djouadi:1998di,Berger:2008cq}. Within this model we will concentrate on the electroweak 
scale parameters that are most relevant for the Higgs and dark matter (DM) sector, that is the 
gaugino, Higgs and third generation squark parameters. We then explore the parameter space of 
the model allowed by flavour constraints, Higgs properties and Higgs searches, direct LHC 
searches for SUSY particles as well as DM constraints (relic density and direct detection).

The collider, flavour and DM constraints on the general pMSSM were explored in several 
publications~\cite{Arbey:2012bp,Dumont:2013npa,Berger:2013zca,Berger:2013zca2,Cahill-Rowley:2014twa,Baer:2014ica, 
Strege:2014ija,Roszkowski:2014iqa,deVries:2015hva}  and the impact of stop searches was also considered~
\cite{Brust:2011tb,Bi:2011ha,He:2011tp,Plehn:2012pr,Kaplan:2012gd, Sekmen:2011cz, AbdusSalam:2012sy, 
Chakraborti:2014gea, Cahill-Rowley:2014twa, Fowlie:2013oua, Roszkowski:2014iqa, Henrot-Versille:2013yma, Strege:2014ija, AbdusSalam:2012ir}. Possible probes to improve the 
bounds on stops and sbottoms with new observables and/or better background reduction have been 
explored in several works~\cite{Plehn:2010st,Bornhauser:2010mw,Bai:2012gs,Han:2012fw,Graesser:2012qy, 
Kilic:2012kw,Ghosh:2012wb,Bai:2013ema,Chakraborty:2013moa,Dutta:2013sta,Buckley:2014fqa,Czakon:2014fka, 
Eifert:2014kea,Cho:2014yma,Ferretti:2015dea, Rolbiecki:2015lsa,Nachman:2013bia} and the difficult case of a compressed spectrum has 
been considered~\cite{Krizka:2012ah,Drees:2012dd,Alves:2012ft,Ghosh:2012ud,Delgado:2012eu,Ghosh:2013qga, 
Belanger:2013oka,Dutta:2013gga,Grober:2014aha,Grober:2015fia,Hikasa:2015lma}. In this work we concentrate 
specifically on the parameter space of the model where a light stop is allowed and we rely on 
simplified model constraints on the stop mass at LHC Run-I. The aim is to understand the 
constraining power of the current search results and demonstrate possible ways in which these 
constraints can be improved. We incorporate recent LHC limits on SUSY particles using 
SModelS~\cite{Kraml:2013mwa,Kraml:2014sna} - a tool that exploits the simplified models results of 
the SUSY searches from the ATLAS and CMS collaborations. Finally, after characterizing the 
remaining parameter space, we discuss the different channels available to further probe light stop 
scenarios in the MSSM. These include LHC SUSY searches, the associated production of stops with light 
Higgs, the monojet search for degenerate stop -- neutralino, searches for heavy Higgses as well as 
direct/indirect DM detection.

This paper is organized as follows. The set-up for the analysis is presented in Section 2 followed 
by a summary of the indirect and direct constraints in Section 3. Our results for the remaining 
available parameter space of the pMSSM with stops below 1.5 TeV are discussed in Section 5 
together with the potential for further probing the model with various collider and astro-particle 
observables. Our conclusions are presented in Section 6.

\section{Analysis set-up}
 Here we have chosen a simplified version of the pMSSM where only the ten 
parameters most  relevant for the Higgs and DM sector  are let to vary. All other parameters 
(squarks of the first and second generation and all sleptons) are fixed to a value large enough  to evade all the LHC constraints. 
The ranges of the values of pMSSM parameters we consider are listed in Table~\ref{scan-range}. 
We performed a flat random scan for values of parameters in these ranges.
Note that sleptons close in mass to the LSP can give an important contribution to DM (co-)annihilation, for example 
this occurs frequently for staus within the constrained MSSM. We ignore this possibility here since the importance of 
coannihilation will be illustrated with third generation squarks, moreover coannihilation with sleptons rely on 
fine-tuning of parameters from unrelated sectors when no assumption is made about the underlying high scale model.

%
\begin{table}[h!]
\begin{center}
\tabulinesep=1.4mm
\hspace*{-7mm}
\begin{tabu}{|l|c|} 
\hline
Parameter                                                          & Scan range            \\
\hline
U(1) gaugino mass parameter: $M_1$                                 &  20 $-$ 2000                   \\
\hline
SU(2) gaugino mass parameter: $M_2$                                &  100 $-$ 2000                  \\
\hline
Ratio of the vacuum expectation values of the two                  &   \multirow{2}{*}{2 $-$ 55}    \\  
Higgs doublets: $\tan\beta$                                      &                                \\
\hline
Higgsino mass parameter: $\mu$                                     & 100 $-$ 3000                   \\
\hline
Pseudo-scalar mass parameter: $m_A$                                & 100 $-$ 2000                  \\
\hline
Stop tri-linear coupling: $A_t$                                    & -5000 $-$ 5000                 \\
\hline
Sbottom tri-linear coupling: $A_b$                                 & -5000 $-$ 5000                 \\
\hline
Mass parameter for the left handed third  generation               &  \multirow{2}{*}{100 $-$ 2000} \\  
squark doublet: $m_{\widetilde{Q}_3}$                              &                                \\
\hline
Mass parameter for the right handed stop:  $m_{\widetilde{U}_3}$   &  100 $-$ 2000                  \\
\hline
Mass parameter for the right handed sbottom: $m_{\widetilde{D}_3}$ &  100 $-$ 2000                  \\
\hline
\end{tabu}
\caption{Scan ranges for the pMSSM parameters, all dimension full parameters are in GeV. The values of all the slepton mass parameters 
as well as $m_{\widetilde{Q}_{1,2}}$, $m_{\widetilde{U}_{1,2}}$, $m_{\widetilde{D}_{1,2}}$ and 
$M_3$ are set to 2 TeV. All the $A$-terms other than $A_t$ and $A_b$ are assumed to vanish.
\label{scan-range}}
\end{center}
\end{table}

We have used SuSpect-2.41~\cite{Djouadi:2002ze} to compute the pMSSM mass spectrum for a given set 
of input parameters. This  includes  two-loop corrections to the Higgs mass and  NLO corrections to  SUSY particle masses. 
In the first step of our scan, we have generated $\sim$ 0.75 Million pMSSM 
points for which the following statements apply, 
\begin{itemize}
\item All the criteria in SuSpect-2.41 for a theoretically valid point are satisfied,
\item The pMSSM spectrum has a neutralino LSP,
%
\item The lightest CP even Higgs boson mass (as computed by SuSpect-2.41) $m_h$ satisfies $118 {\rm GeV}< m_h < 130 {\rm GeV}$,  
\item The lightest stop and the LSP satisfy $\mstl < 1500$ GeV and 
$\mlsp < 800$ GeV respectively.
\end{itemize}

Once the above step is done, we assess the impact of Higgs boson signal strengths, flavour violating 
observables, DM relic density and direct detection cross section, LEP data and finally, 
direct  SUSY searches at the LHC on the selected set of  $\sim$ 0.75 Million pMSSM points. 

Note that we have checked the vacuum stability and the absence of charge and colour breaking minima via SuSpect. A more refined analysis along the lines of \cite{Chowdhury:2013dka,Camargo-Molina:2014pwa,Blinov:2013fta} might improve the constraints in the large $A_t$ region.

\section{Indirect constraints}
\begin{itemize}
\item \underline{Higgs data} :  In order to study the compatibility of the pMSSM models with Higgs 
data we use HiggsBounds-4.1.0~\cite{Bechtle:2008jh,Bechtle:2011sb} and HiggsSignals-1.1.0~\cite{Bechtle:2013xfa} 
which are linked to FeynHiggs for the computation of Higgs mass and signal 
strengths. As far as the theoretical uncertainty in the Higgs mass calculation is concerned, 
we use the estimate given by FeynHiggs-2.10.0~\cite{Heinemeyer:1998yj}. 
While calculating the $p$-value in HiggsSignals, we set the number of free model parameters $N_p$ = 10. 
The $p$-value is required to be more than 0.05 for an allowed parameter point. Note that the  range for 
the Higgs mass assumed in our preselection will automatically be reduced by imposing these constraints.
\item \underline{Flavour data} :  
Flavour physics has played a crucial role in the construction of the SM as well as constraining NP 
beyond the SM. The flavour structure of the SM is indeed very special, and any generic NP model suffers 
from large flavour violations in contradiction with the wealth of data from B-factories and recently 
also from LHCb. In the pMSSM, because of the degeneracy of the first two squark generations the flavour 
violation involving the first two generation of fermions is mild. Moreover, as we have set the masses 
of the first two generation of squarks to 2 TeV (which is allowed by direct searches at the LHC), 
their contributions decouple. Thus, the flavour constraints mainly arise from processes involving the 
third generation of quarks, for example, decays involving $b \to s$ transition
\footnote{Interestingly, the LHCb collaboration has reported hints 
of NP in some of the $B$ meson decay modes involving quark level $b \to s$ 
transitions~\cite{Aaij:2013qta,Aaij:2014ora}. Although, 
there are pending issues with the reliability of the theoretical SM predictions and firm claims 
of the existence of NP can not be made yet, several NP explanations of these ``deviations'' have 
been proposed~\cite{Descotes-Genon:2013wba,Gauld:2013qba,Datta:2013kja,Altmannshofer:2014cfa,
Ghosh:2014awa}. 
Unfortunately, a NP explanation within the MSSM seems unlikely~\cite{Altmannshofer:2014rta}.}. 
In particular, the flavour changing $B$-meson decays e.g., the radiative decay 
$\br(B_d \to X_s  \gamma)$ and the fully leptonic decay $\br(B_s \to \mu^+  \mu^-)$ 
are known to place important constraints on the MSSM parameter space. 
In this work we have used the following limits for these branching ratios.
\bea
2.78 \times 10^{-4}  \leq & \br(B_d \to X_s  \gamma) & \leq 4.08 \times 10^{-4} \, ,  \\
1.43 \times 10^{-9}  \leq & \br(B_s \to \mu^+  \mu^-)& \leq 4.37 \times 10^{-9} \, .
\eea
In addition, we have also imposed the following limits on the 
branching ratio of $B_d \to \mu^+  \mu^-$,
\bea
0.79 \times 10^{-10} \leq & \br(B_d \to \mu^+  \mu^-)& \leq 6.80 \times 10^{-10} \, \,.
\eea
In the numerical analysis, we have used the public code SuperIso-3.3~\cite{Mahmoudi:2008tp} 
to compute these branching ratios.

Note that we have not considered a few other potentially important observables such as the 
anomalous magnetic moment of the muon, $(g-2)_\mu$, the
branching ratios of the two body leptonic decay of the $B$-meson $\br(B \to \tau \nu)$
and the three body semileptonic decay $\br(B \to D(D^*)\tau \nu)$. As far as $(g-2)_\mu$
is concerned, the SUSY contribution needs to be at least $1.0 \times 10^{-9}$ in order to be
consistent with the measured value at the 2$\sigma$ level~\cite{Bennett:2006fi,Gnendiger:2013pva}.
This requires the existence of light sleptons and electroweak gauginos  which can be easily achieved in the  pMSSM~\cite{Chakraborti:2014gea}.  
\comment{\tr{However we did not want to confine the analysis to the region compatible with the $(g-2)_\mu$ 
constraint until more precise measurements and theoretical SM predictions are available 
\cite{Knecht:2014sea}.}}
 We set the slepton mass parameters to a high value for simplicity.

The leptonic decay $B \to \tau \nu$ which has tree level SUSY contribution from the charged 
Higgs exchange diagrams is also known to provide stringent constraints on the SUSY parameter 
space~\cite{Bhattacherjee:2010ju}. Interestingly, the most recent measurement by the Belle
collaboration has brought down  this branching ratio to a much smaller value than earlier 
measured~\cite{Adachi:2012mm}, thus relaxing the tension with the SM prediction. All our points satisfy this constraint 
since many other constraints force the charged Higgs to be rather heavy any way.

Let us now discuss about the three body semileptonic decay $\br(B \to D \tau \nu)$ and 
$\br(B \to D^*\tau \nu)$. The BaBar collaboration measured the two quantities~\cite{Lees:2012xj}
$$R(D) = \frac{\br(\bar{B} \to D \tau^{-} \bar{\nu_\tau})}
{\br(\bar{B} \to D \ell^{-} \bar{\nu_\ell})} \, \, \textnormal{and} \, \,
R(D^*) = \frac{\br(\bar{B} \to D^* \tau^{-} \bar{\nu_\tau})}
{\br(\bar{B} \to D^* \ell^{-} \bar{\nu_\ell})},$$
and reported a 3.4$\sigma$ deviation from the SM when the two measurements are taken together. This 
result motivated a number a phenomenological studies both in the context of specific 
models~\cite{Fajfer:2012jt} as well as model independent approaches~\cite{Fajfer:2012vx,Datta:2012qk}. 
The BaBar collaboration itself ruled out a Type-II Two Higgs Doublet Model (THDM) at 99.8\% 
confidence level for any value of $\tan\beta/m_{H^\pm}$ based on this data. The same would apply 
to the Higgs sector of the MSSM which is also a  Type-II THDM at the tree level. However, the  
existence of two neutrinos in the final state of these decays makes their measurements quite 
challenging experimentally and a confirmation of these results by another independent experiment 
(e.g., Belle II) is awaited.

\item \underline{Dark matter relic density and direct detection} : 
The DM relic density has been measured precisely by PLANCK, 
$\Omega h^2=0.1192\pm 0.00024$ at 68\% CL~\cite{Ade:2013zuv}. 
We impose the following upper bound 
\bea
\Omega h^2 \leq 0.129
\eea
which corresponds to the measured value after adding a 10\% theoretical uncertainty - this 
number is a rough estimate of uncertainties that can arise for example from one-loop corrections 
to DM annihilation cross section~\cite{Baro:2007em,Baro:2009na,Boudjema:2011ig,Harz:2012fz,Harz:2014tma} 
\footnote{A more precise value for the relic density has been released recently by 
PLANCK~\cite{Planck:2015xua}, 
this however has no impact on the results presented here since the theoretical uncertainty we 
assume is dominant.}. We impose only the upper bound to allow for the possibility that the 
neutralino is only a fraction of the DM.
\comment{ or that there is some regeneration mechanism 
that comes into play for example from decay of moduli fields re-injecting neutralinos after the 
freeze-out~\cite{Moroi:1999zb,Hall:2009bx,Arcadi:2011ev}.}

We also impose  the mass dependent upper bound on the WIMP direct detection cross section obtained 
by the LUX experiment. For this,  we have fitted the LUX upper bound~\cite{Akerib:2013tjd}
to an analytic formula which is given by, 
\bea
\label{lux-formula}
\rm \log_{10} \sigma_{SI}^{LUX} = && \dfrac{7.029}{(\rm \log_{10} m_{WIMP})^2} - 
\dfrac{7.161}{\rm \log_{10} m_{WIMP}} - 8.569  \nonumber \\ 
&& + \, 0.755 \rm \log_{10} m_{WIMP} - 0.003 \, (\rm \log_{10} m_{WIMP})^2 \,\, .
\eea
Based on this we  apply the following constraint,
\beq
\label{relic-limits}
\rm \sigma_{SI} <  \rm \xi \sigma_{SI}^{LUX} 
\eeq
where

\beq
\label{relic-limits}
\xi=
\left\{ 
\begin{array}{lll}
1  &         \, \, \rm if \; \;               &   0.1103 < \Omega h^2 < 0.1289 \,\, ,  \\
                               \\                         
\dfrac{0.1196}{\Omega h^2}                        &        \, \, \rm if \; \;                     
& \Omega h^2 < 0.1103 \, \, .         
\end{array} \right.
\eeq
\end{itemize}
That is, if the relic density computed assuming the standard cosmological scenario falls below 
the PLANCK range, we consider that the neutralino constitutes only a fraction of the DM 
and explicitly ignore the possibility of regenerating DM although we will comment on 
this assumption in section~\ref{sec:DM}. Moreover we do not make any assumptions about what 
would constitute the rest of the DM. Note that DM can also be searched for by indirect  detection. 
In our analysis we have neither imposed the constraints from FermiLAT on photons~\cite{Ackermann:2015zua}, 
from PAMELA on antiprotons~\cite{Adriani:2010rc} as well as the preliminary limits from AMS on 
antiprotons~\cite{AMS-antiproton},  nor have we made any attempt to explain the anomalies observed by 
PAMELA~\cite{Adriani:2008zr} and AMS~\cite{Aguilar:2013qda} on the positron spectrum.  
We do however, briefly discuss the impact of indirect searches in section~\ref{sec:DM}. We use 
micrOMEGAs-3.5.5~\cite{Belanger:2013oya} to calculate the DM relic density as well as 
the direct and indirect detection cross sections.

In the MSSM, it is well known that the composition of the neutralino is crucial for 
determining the DM properties. For neutralino annihilation to be efficient enough to have 
$\Omega h^2 \le 0.129$ requires either a LSP with a significant Higgsino or wino fraction or 
special tuning of parameters. In fact a dominantly Higgsino/wino LSP with a mass in the range 
80~GeV to 1 $-$ 2 TeV typically leads to $\Omega h^2 < \Omega_{\rm PLANCK} h^2$ because of efficient 
annihilation into W pairs. A mixed state with some bino component is therefore preferred. Direct 
detection cross section on the other hand is large for a mixed gaugino/Higgsino LSP, in 
particular for the one that leads to the exact range of the relic density determined by PLANCK. 
The only possibility for such mixed neutralino would be to lie above the TeV scale where the 
direct detection limits are weaker, we do not consider these masses since we want to highlight 
the SUSY spectrum below the TeV scale. Both cases with pure wino or pure Higgsino DM easily 
evade the direct detection constraints although the DM relic density cannot be entirely 
explained by neutralinos.  Another possibility which allows also for a dominantly bino LSP 
consists in adjusting parameters such that $m_{\widetilde{\chi}}\approx m_Z/2$ or $m_h/2$ or $m_H/2$ thus 
providing a 
resonant enhancement of the cross section or having $m_{\widetilde{\chi}} \approx m_{sfermion}$. The 
contribution of coannihilation channels then reduces the relic density. The dominantly bino LSP 
is only allowed for light sfermions or when the mass is such that one can benefit from 
annihilation through a resonance in s-channel.  Thus we expect to find a large number of 
scenarios with dominantly Higgsino or wino LSPs.

\section{Direct search constraints}

\begin{itemize}
\item \underline{LEP limits} :

The generic limits from LEP are obtained directly from micrOMEGAs and mainly exclude  charged 
particles. The lower limit on chargino is 103 GeV while those on sleptons, in particular staus are 
slightly weaker. In addition we have also imposed an upper limit on the Z invisible width, 
$\Gamma_Z < $ 2 MeV as well as constraints on neutralinos from $\sigma (e^+e^-\to 
\neutl\widetilde{\chi}_i^0)< 0.1$ pb where the heavy neutralino, $\widetilde{\chi}_i^0$, decays mostly 
into hadrons and a LSP~\cite{Abbiendi:2003sc}. Such analyses constrain the very light neutralino 
LSP region, which is also strongly constrained by the upper bound on the DM relic density.

Finally we impose the condition that all particles decay promptly ($c\tau < 0.05$ m). A strict 
requirement is that charged particles are not long-lived at the cosmological scale, we impose 
this more restrictive criteria because the LHC limits that we implement below assume prompt 
decays leading to Missing Transverse Energy (MET) in the final state. 

\item \underline{LHC limits} : Searches for SUSY at the LHC form an important ingredient to assess 
the viability of the scenario under consideration. Here, we describe our procedure for 
evaluating the LHC constraints. We use \smodels~\cite{Kraml:2013mwa,Kraml:2014sna}, a tool 
designed to evaluate the LHC constraints on NP using simplified 
model spectra (SMS) results. \smodels is designed to decompose the signal of any arbitrary NP
spectrum with a $\mathbb{Z}_2$ symmetry into simplified model topologies and test it against the 
existing LHC bounds in the SMS context. The input to \smodels can either be an SLHA 
file~\cite{Skands:2003cj} containing the SUSY production cross sections $\sigma$ and the branching 
ratios for the SUSY decays, $\br$ or a LHE file~\cite{Alwall:2006yp}. For this work we 
used the SLHA input containing $\sigma$ and \br. The format for writing the production 
cross sections is 
specified in~\cite{SLHA-xsections}. The production cross sections are computed using 
Pythia6.4.27~\cite{Sjostrand:2006za} and NLL-fast-2.1 
~\cite{nllfast,Beenakker:1996ch,Kulesza:2008jb,Kulesza:2009kq,Beenakker:2009ha,Beenakker:2011fu, 
Beenakker:1997ut,Beenakker:2010nq}. Given the information on the $\sigma$ and \br, \smodels 
computes \sigmaXBF for each possible decay of SUSY particles. The information of relevance to 
check the results against the LHC limits is the mass vector of the SUSY particles, the SM decay 
products and the $\sigmaXBF$ of the resulting topologies. A topology resulting from such SLHA 
decomposition is considered if the $\sigmaXBF > \sigma_{\rm cut}$, with $\sigma_{\rm cut}$ set to 
0.01 fb.

When dealing with an arbitrary NP spectrum, care must be taken to identify regions of compressed 
spectra as the decay products in such cases are not detected. \smodels ignores such soft decays 
when the mass gap between the mother and the daughter particles is less than the user defined 
minimum mass gap, here we take 5 GeV. 
The resulting \sigmaXBF for various SMS topologies were tested against ATLAS SMS 
interpretation for searches~\cite{Aad:2013wta,Aad:2013ija,Aad:2014nua,Aad:2014qaa,Aad:2014vma, 
Aad:2014wea,Aad:2014yka,Aad:2014kra,ATLAS-CONF-2012-105,ATLAS-CONF-2013-007,ATLAS-CONF-2013-024, 
ATLAS-CONF-2013-061,ATLAS-CONF-2013-065} and CMS searches~\cite{Chatrchyan:2013wxa,Chatrchyan:2013lya,
Chatrchyan:2014aea,Khachatryan:2014qwa, 
Chatrchyan:2013iqa,Chatrchyan:2013xna,Chatrchyan:2014lfa,Chatrchyan:2013fea,CMS-PAS-SUS-13-008, 
CMS-PAS-SUS-13-016,CMS-PAS-SUS-13-018,CMS-PAS-SUS-13-019,CMS-PAS-SUS-14-011}. 
\end{itemize}

\section{Results}

We start our discussion with Table~\ref{cut-flow} where the impact of the various experimental data 
on the pMSSM parameter space is shown. Since we are randomly scanning over 10 uncorrelated parameters, 
it is important to generate enough points to populate all dimensions. The table shows the effect of each successive 
experimental constraint on a well populated flat scan over 10 dimensional parameter space. The number of pMSSM 
points which survive after each successive constraint  is presented in the second column. 
Although no statistical meaning can be attached to these numbers, Table~\ref{cut-flow} and the accompanying 
discussion helps to get an understanding of how the various type of data constrain the MSSM parameter space.
It can be seen that out of the $\sim$ three quarter of a million 
points that were generated, only 60\% of them successfully satisfy the observed Higgs boson mass, 
signal strength data and other Higgs boson searches (which are implemented in HiggsBounds-4.1.0 and 
HiggsSignals-1.1.0).  
\begin{table}[h!]
\begin{center}
\tabulinesep=1.4mm
\hspace*{-7mm}
\begin{tabu}{|l|c|} 
\hline
Constraints                                                     &   No. of models        \\
\hline
Theory {\bf +} neutral LSP {\bf +} &\\ $118 {\rm GeV} < m_h < 130 {\rm GeV}$  
{\bf +} $\mstl < 1500 {\rm GeV}$ 
{\bf +} $\mlsp < 800 {\rm GeV}$                        &   741605               \\ 
\hline
\multicolumn{2}{|c|}{{\bf Indirect bounds}}                                              \\
\hline
{\bf +} HiggsBounds + HiggsSignals                              &   435021               \\ 
\hline
{\bf +} $\br(B_d \to X_s  \gamma)$                               &   211313               \\ 
\hline
{\bf +} $\br(B_{s,\,d} \to \mu^+  \mu^-)$                        &   177961               \\ 
\hline
{\bf +} $\Omega h^2_{<}$ + $\rm \sigma_{SI}^{LUX}$              &   111167               \\ 
\hline
\multicolumn{2}{|c|}{{\bf Direct bounds}}                                                \\
\hline
{\bf +} LEP                                                     &   59425                \\  
\hline
{\bf +} Long-lived chargino                                     &   30754                \\ 
\hline
{\bf +} LHC                                                     &   29266                \\ 
\hline
\end{tabu}
\caption{Number of surviving pMSSM models after each cut. 
\label{cut-flow}}
\end{center}
\end{table}
The severe impact of the measured branching ratio of $B_d \to X_s \gamma$ is also clear from the 
table. Out of all the models which satisfied Higgs data, only about 50\% survive after the 
consistency with $B_d \to X_s \gamma$ branching ratio is imposed. It is worth mentioning that 
there is some amount of tension between the Higgs mass and the branching ratio of $B_d \to X_s 
\gamma$ in the MSSM. The latter gets contributions from chargino loops as well as charged Higgs 
loops. As the charged Higgs loop interferes constructively with the SM, its contribution is 
always positive. The chargino contribution on the other hand can be both constructive or 
destructive. For example, the stop$-$Higgsino contribution to the amplitude is proportional to 
$m_t^2 \mu A_t \tan\beta/m_{\widetilde{t}}^4$ and hence, this contribution is enhanced for a 
large $A_t$. On the other hand, consistency with the measured value of the Higgs mass requires a 
large $A_t$ if a light $\stopl$ is desired.

The DM constraints also reduce significantly the number of allowed points. In 
particular when the LSP is lighter than the W the only allowed points are near $m_Z/2$ or 
$m_h/2$ corresponding to the annihilation through a Z or Higgs resonance. For heavier LSP masses 
the relic density upper bound basically selects LSP with large Higgsino or wino LSP, barring 
special configurations where coannihilations are important. Furthermore the mixed 
bino(wino)/Higgsino can be in conflict with the LUX direct detection bound as mentioned above.  
We therefore expect that the LSP will most of the time be dominantly Higgsino or wino which in 
both cases implies that it is almost degenerate with the chargino. Such degenerate chargino - 
neutralino lead to long lived charginos, leading to charged tracks at the LHC. As explained 
before, these scenarios are not considered in the present study and hence nearly half the points 
are discarded. Similarly the LEP limits on charged particles also rule out nearly half of the 
allowed points. Finally, LHC limits from the 8 TeV run rule out about 5\% of the allowed points. 
Below we discuss in more details the impact of LHC constraints from SUSY searches as obtained 
with \smodels.

\begin{figure}[h!]
\begin{center}
\begin{tabular}{ccc}
\hspace{-12mm}\includegraphics[scale=0.4]{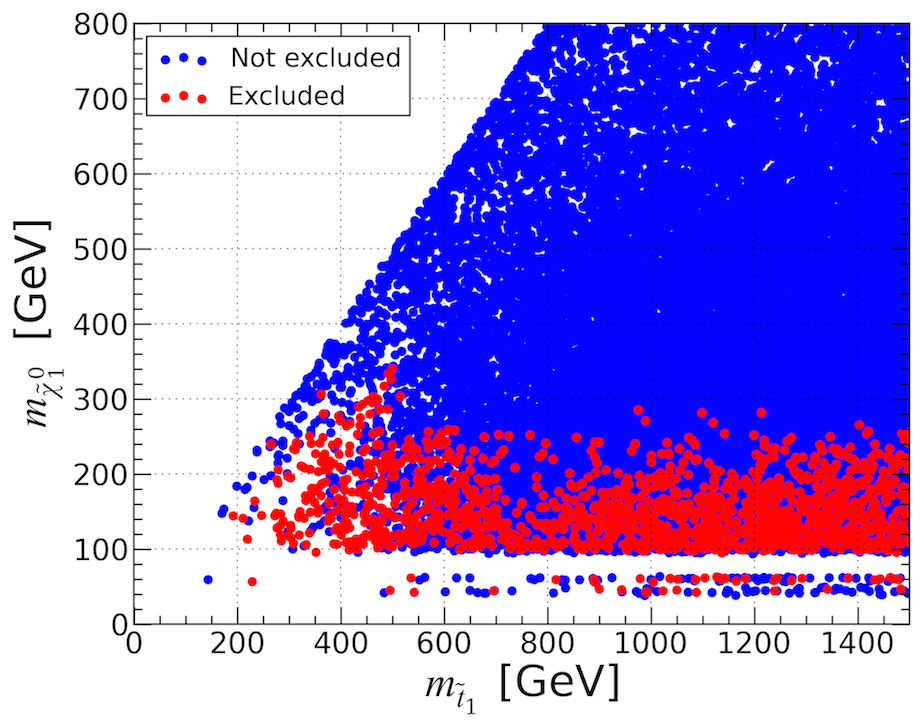}&
\includegraphics[scale=0.4]{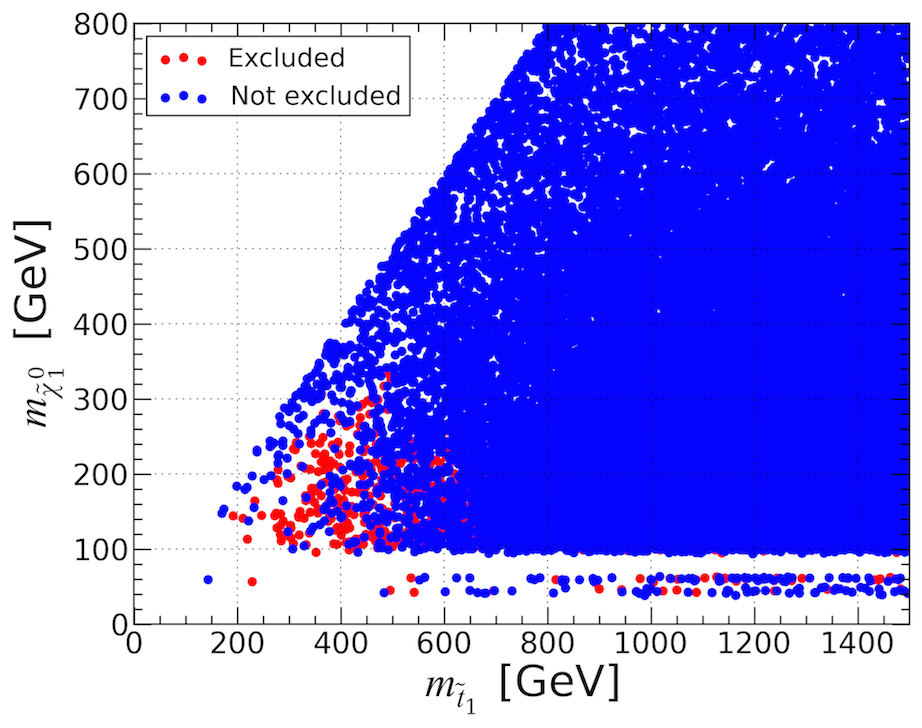}
\end{tabular}
\caption[]{Summary of the allowed and excluded points by \smodels. In the left panel, 
the excluded points are displayed on top of the allowed points while in the right panel the 
allowed points are on top of the excluded points. 
\label{fig:stop_smodels_summary}}
\end{center}
\end{figure}

In Fig.~\ref{fig:stop_smodels_summary} we plot the allowed and excluded points after applying 
\smodels. In the left panel we plot the excluded points on top of the allowed points while in 
the right panel, the plotting order is inverted.  Clearly many points with light stop masses are 
not excluded by SMS results as implemented in \smodels-1.0.3. It is possible to exclude many 
points up to the maximum stop mass considered (1500GeV). However, no SMS result has reach for 
LSP masses greater than 300 GeV and therefore the region with higher LSP masses remains 
unconstrained.

The right panel shows that different configurations of the MSSM spectra can evade the SMS 
results, thus allowing very light stop masses $-$ even below 200 GeV. Despite the fact that 
\smodels combines topologies with the same final states and similar mass vectors, the main reason 
for the allowed region is that the SMS results obtained by LHC collaborations and used by 
\smodels assume a 100\% branching ratio for the decay under consideration while in the MSSM 
branching ratios are often below 100\%. 
For  example consider a point in our scan with $\mstl \approx 380$ GeV and the rest of the spectra too heavy to contribute to limit setting.
The dominant branching ratio for the stop is into $b\tilde\chi^+$ (77\%), however this channel cannot be exploited as the chargino decays into a virtual W and the LSP,
 a channel which is not implemented in SModelS. Therefore only the  channel  $t\tilde\chi_1^0$  (with a BR $\approx 0.13$) can be used to constrain the stop. This point gives $\sigmaXBF \approx 8.29$ fb, whereas the experimental upper limit in CMS-SUS-13-011 derived assuming a 100\% $\mathcal{B}$ is 438 fb. Hence this point does not get excluded by \smodels. The right panel of Fig. 1  thus justifies the possibility that a light stop is consistent with the current LHC SUSY searches as well as the observed Higgs properties and the heavy Higgs searches, along with the DM direct detection limits.
\begin{figure}[h!]
\begin{center}
\begin{tabular}{ccc}
\hspace{-12mm}\includegraphics[scale=0.5]{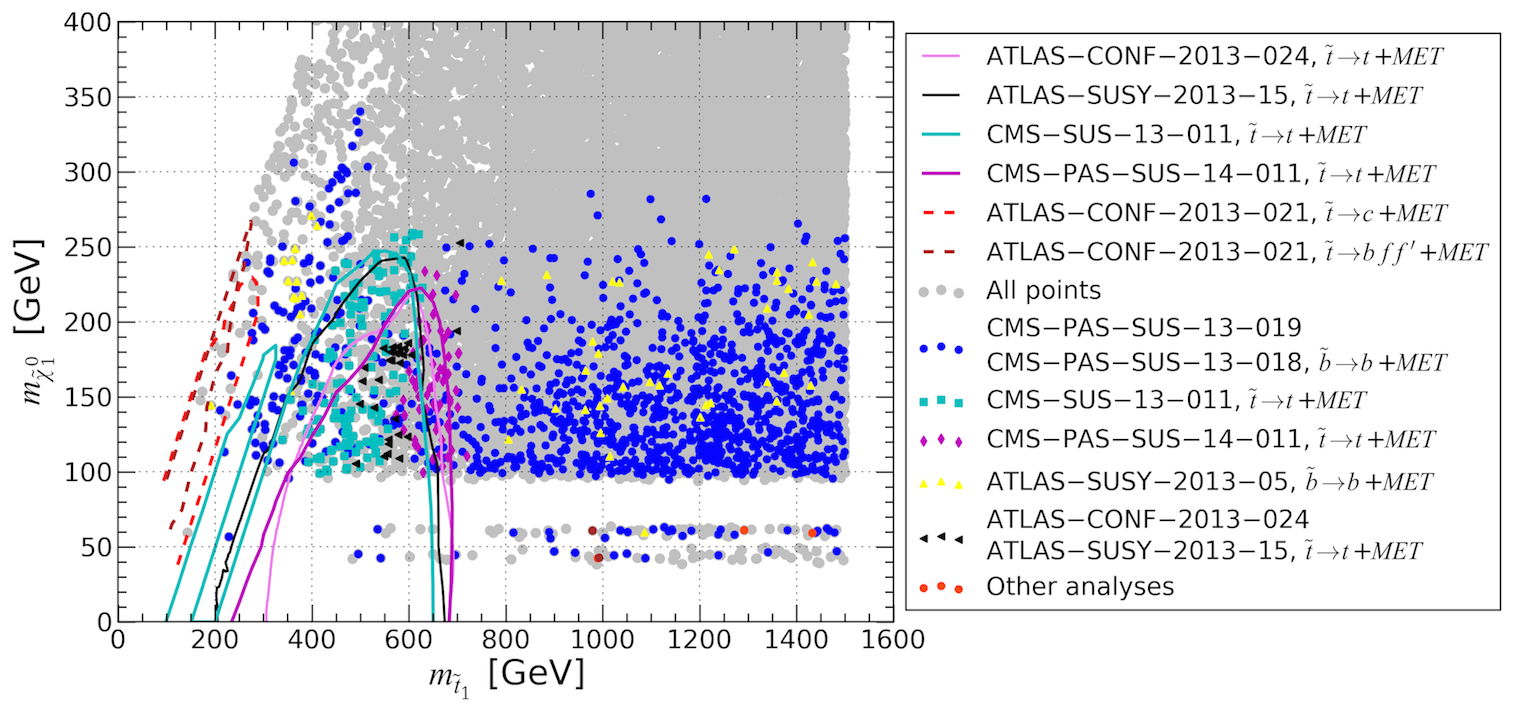}\\
\hspace{-12mm}\includegraphics[scale=0.5]{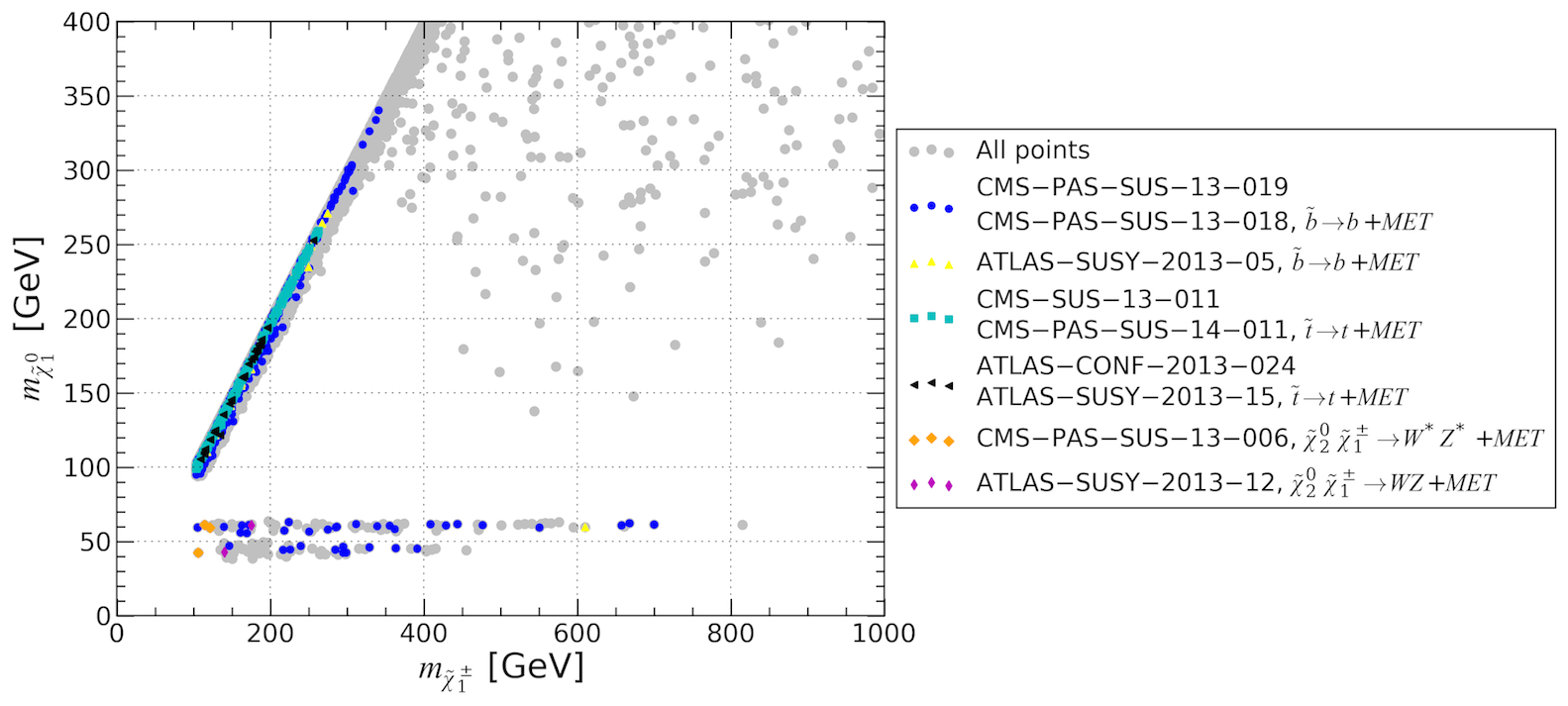}
\end{tabular}
\caption[]{Breakdown of the most constraining analysis for the points excluded by \smodels  
in the $\mstl - \mlsp$ mass plane (upper panel) and in the  $\mc1 - \mlsp$ mass plane (lower panel). 
For comparison, exclusion curves from various ATLAS and CMS stop searches are overlaid (solid lines). 
The dashed line shows exclusions arising from ATLAS searche for stop decays to $c\neutl$ and 
to $b\neutl f f'$.\label{fig:smodels_summary_breakdown}}
\end{center}
\end{figure}
 
Figure~\ref{fig:smodels_summary_breakdown} shows the breakdown of the most constraining analysis 
for the excluded points in the stop -- LSP mass plane and chargino -- LSP plane. For each of the 
excluded points, we select the most constraining analysis~\footnote{The most constraining 
analysis is defined as the analysis which leads to the largest ratio of the theory cross section to 
the experimental upper limit.}. It is possible that a point is excluded by more than one 
searches at the LHC. We overlay the exclusion lines obtained by ATLAS and CMS from stop 
searches to guide the eye. Indeed we see that most of the points excluded by the 
$\tilde{t}\to t\neutl$ searches fall within the corresponding exclusion contours 
while higher stop masses are in fact excluded by constraints coming from sbottom searches. There 
are also significant constraints arising from sbottom searches for the light stop masses in the 
regime where $\tilde{t}\to t\neutl$ is kinematically forbidden while 
$\tilde{b}\to b\neutl$ is allowed. This kinematic configuration is common when 
$\stopl,\sbottom1$ are left-handed, thus sbottom searches provide indirect constraint on 
the light stop scenarios otherwise elusive at the LHC. For the kinematic edge where $m_{\tilde 
t} - \mlsp < m_{t}$, four body decays of stops as well as decays via 
$c\neutl$ are also utilized at the LHC. Results from ATLAS~\cite{Aad:2014nra} (for four 
body stop decay as well as decay via charm) and CMS~\cite{CMS-PAS-SUS-13-009} (stop decays via 
charm) searches are available. The ATLAS search yields stronger limits. Unfortunately, the cross section
95\% C.L. observed upper limit map on the \sigmaXBF is not available in~\cite{Aad:2014nra}. This 
search hence could not be included in \smodels and was not used in this study. The red dashed 
exclusion lines obtained from the ATLAS searches ~\cite{Aad:2014nra} are overlaid in 
Fig.~\ref{fig:smodels_summary_breakdown} (upper panel) for comparison. It can be seen that they 
do not cover a large region of parameter space, hence do not affect our conclusions drastically.

Figure~\ref{fig:smodels_summary_breakdown} (lower panel) shows the exclusions in the $\charginol 
- \neutl$ plane, three distinct branches can be seen. The points along $\mlsp 
\approx 45$ GeV or $\mlsp \approx 60$ GeV correspond to mostly bino-like $\neutl$, while 
the points along the diagonal represent dominantly Higgsino or wino $\neutl$ .
The figure illustrates that most of the times, the chargino is nearly mass degenerate with the 
LSP and either decays invisibly or via an off-shell $W$ to LSP, thus evading the SMS limits 
which do not include either of these channels. The current searches for $\neuth \charginol$ 
decaying to WZ + MET therefore exclude only a few points in this 
parameter space. These are located along the bino branches of the plot. Furthermore, along the 
diagonal lines where the chargino decay is invisible, the searches for stop and sbottom pair 
production with direct decays to LSP or Next-to-Lightest Supersymmetric Particle (NLSP) contribute 
the most to the exclusion.

\begin{figure}[h!]
\begin{center}
\begin{tabular}{ccc}
\includegraphics[scale=0.38]{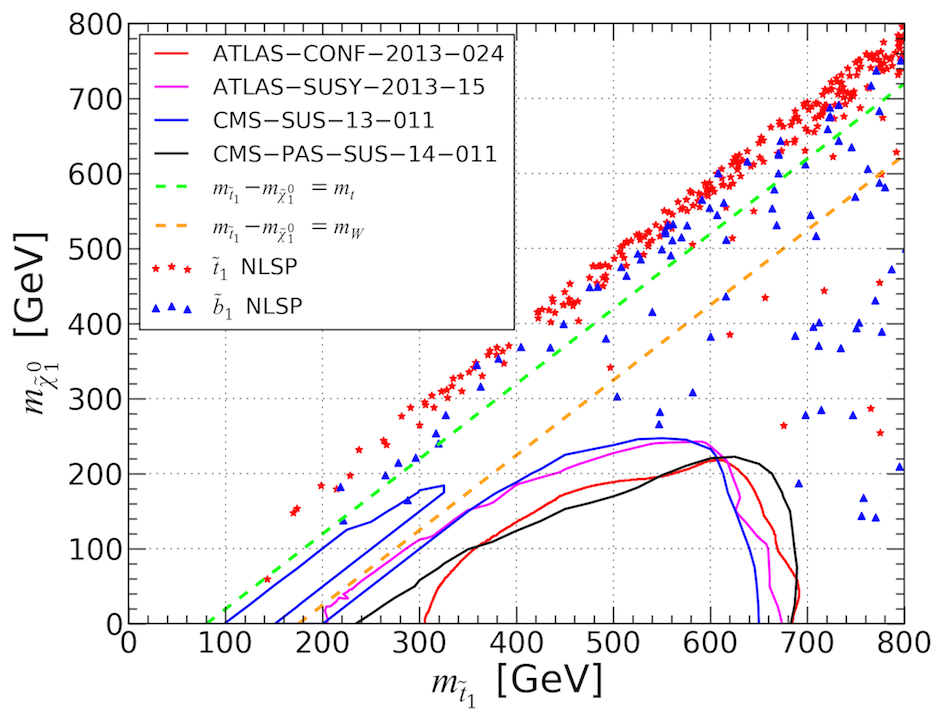}&
\includegraphics[scale=0.38]{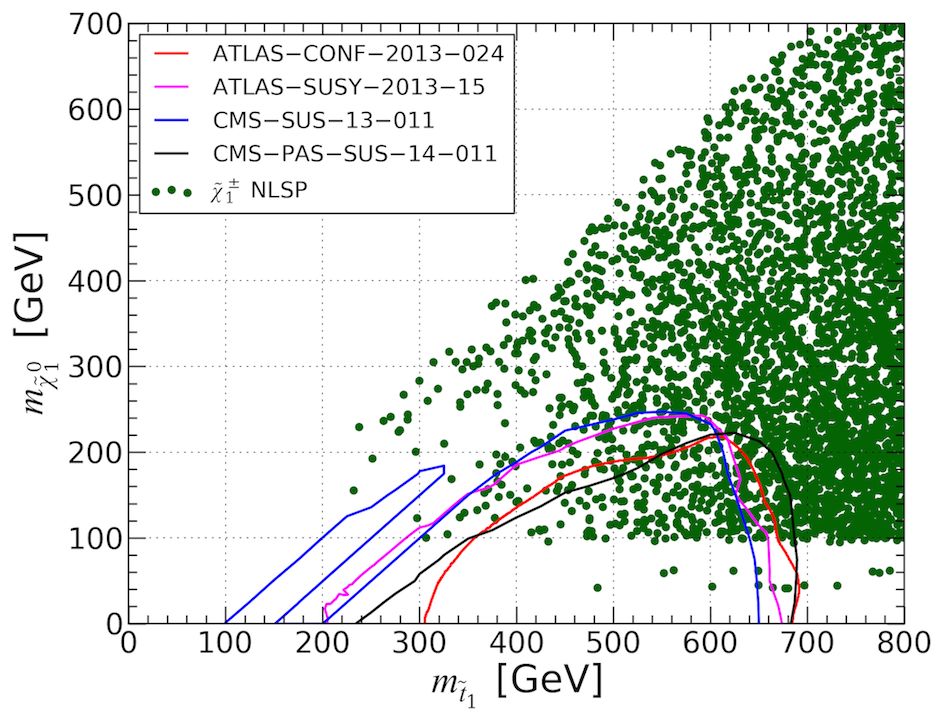}
\end{tabular}
\caption[]{Allowed points in the $\mstl-\mlsp$ plane when the NLSP is $\mstl$ or $\msbl$ (left panel), 
$\mc1$ (right panel). 
The exclusions curves from the analyses of the stop searches used in testing the points in the 
$\mstl - \mlsp$ plane. The lines with different colors  and styles correspond to different SMS
results for direct stop pair production decaying to $t\bar t +\textrm{MET}$.
\label{fig:NLSP_allowed_breakdown}}
\end{center}
\end{figure}

In Fig.~\ref{fig:NLSP_allowed_breakdown} we plot the allowed points  in the $\mstl - \mlsp$ plane. 
We separate the total number of allowed points in three categories 
i) pMSSM points which have the lightest stop as the  NSLP (red stars)
ii) lightest sbottom as the NLSP (blue triangles) iii) lightest chargino as the NLSP (green circles). 
We have shown these three set of points separately to emphasize that most of the allowed parameter points 
where $\stopl$ is the NLSP (the left panel in Fig.~\ref{fig:NLSP_allowed_breakdown}) lie close 
to the $\mstl = \mlsp$ line. The requirements on the relic density leading to the stop 
co-annihilation region is responsible for this strip. In these kind of scenarios, it is 
difficult to constrain these points via direct stop searches. The right panel shows 
that most of the points belong to the category of chargino NSLP as argued above. 
In such cases, the stop has cascade decays via chargino or heavier neutralinos, thus reducing the 
branching ratio for each mode and leading to weaker exclusions than expected from SMS results.
Moreover only results where the chargino decays through a real W are included in SmodelS.

We have demonstrated the exclusions obtained with the help of SMS results. However, it might 
be possible to obtain stronger exclusions by means of recasting an analysis, for example using 
the approach in~\cite{Kim:2015wza,Dumont:2014tja,Conte:2014zja}. This is clearly beyond the 
scope of this study. It is  worth noting that allowing for light  sleptons in the scan might 
lead to further exclusions  driven by  chargino - neutralino decays via intermediate sleptons.
Furthermore, there are SMS interpretations for the decays of heavier stop e.g. stop searches 
with final states involving Higgs or Z boson~\cite{Khachatryan:2014doa}. As will be 
demonstrated in section~\ref{sec:heavystop} due to the constraint on the Higgs mass the 
$\stoph$ is always heavier than ~700 GeV, where these searches currently  do not have 
sensitivity. 

\section{Probing light stop scenarios at the LHC}

From the results just presented it should be clear that the light stop
scenarios in the pMSSM offer a variety of signatures at colliders. Note
that this includes not only those  from direct stop production but also
from other light super particles. In particular, a light left-handed
stop means, quite often, a light left-handed sbottom as well. Thus, in this section
we  first investigate the main signatures that could not be constrained
by SModelS and suggest additional topologies which may be pursued at
LHC14. Here we concentrate on final states produced by stop and sbottom
decays and show missing topologies with large cross-section. We further
also present missing topologies that arise from the electroweak sector
for our allowed MSSM points. Then we examine other potential signatures
from light stop associated production with an extra jet or a Higgs.
Furthermore, we examine possible final states resulting from the decay
of the heavier stop and we also discuss aspects of Heavy Higgs
phenomenology for the set of MSSM points  which are allowed in our light
stop scenario.

\subsection{Improving simplified models interpretations at 8 TeV }

\begin{figure}[h!]
\begin{center}
\begin{tabular}{l}
\includegraphics[scale=0.105]{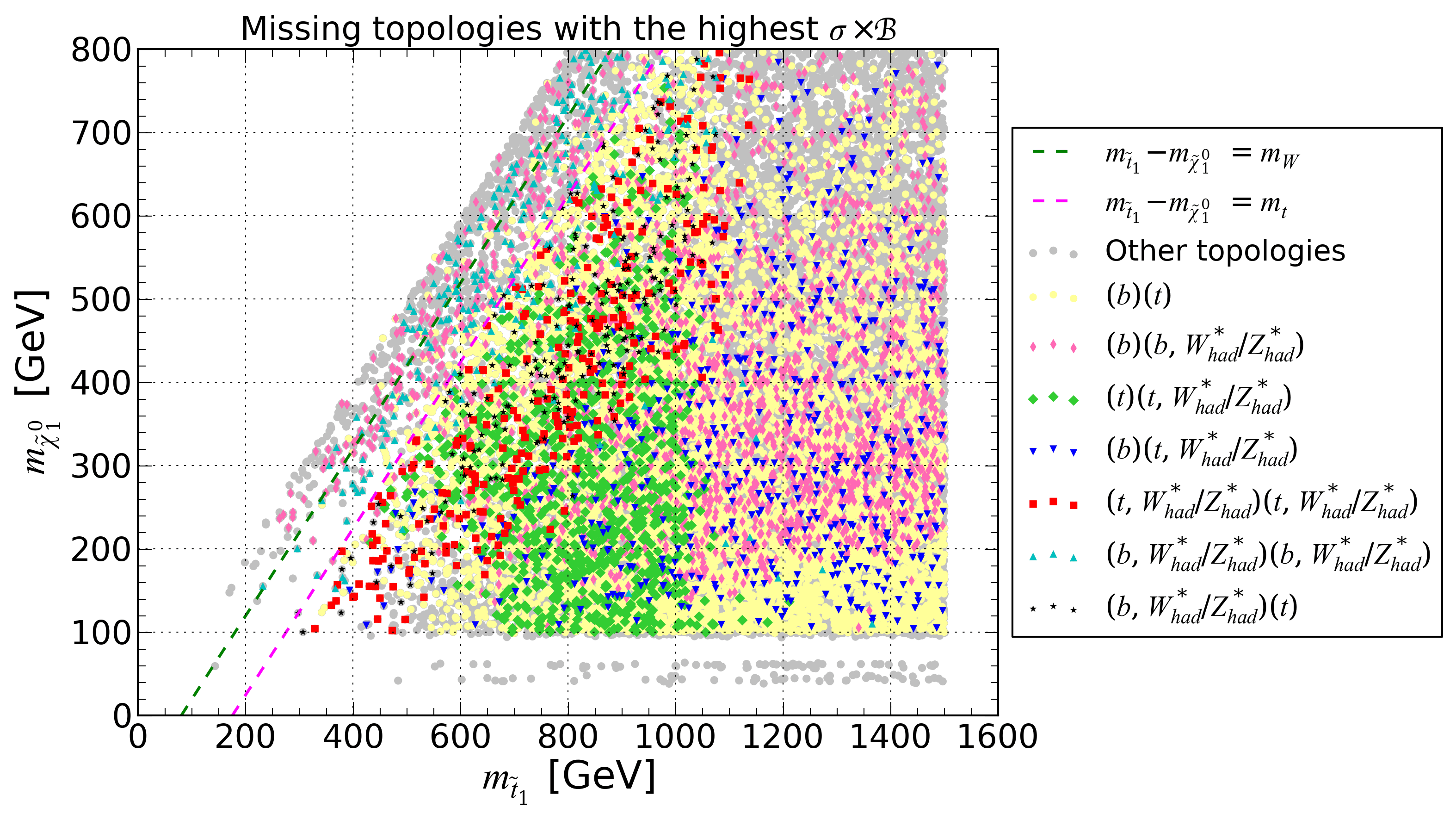}\\
\includegraphics[scale=0.105]{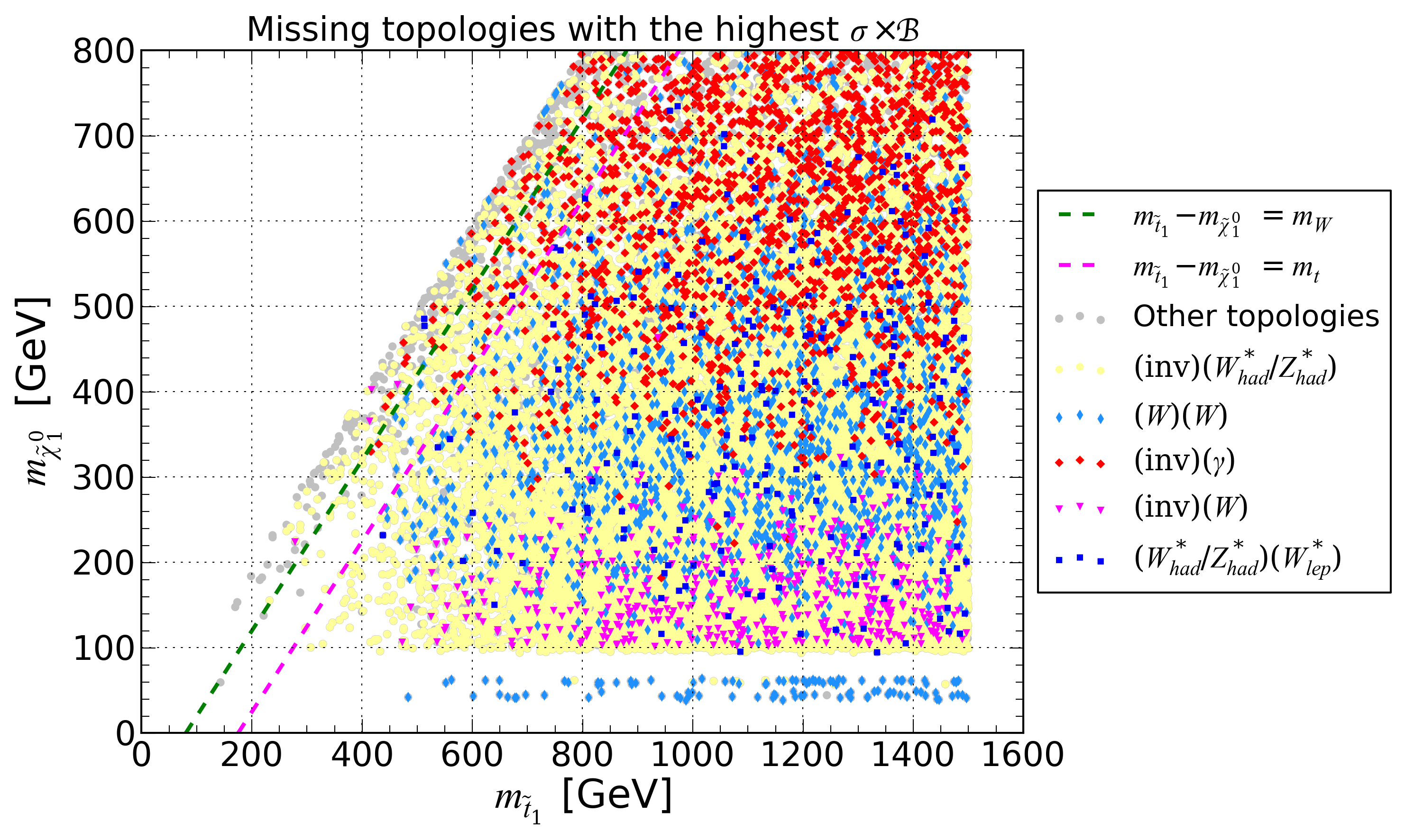}
\end{tabular}
\caption[]{Most dominant missing topologies in the stop - LSP mass plane. For each non-excluded point, 
the SMS topology leading to the highest \sigmaXBF is chosen. The seven most frequently occurring 
topologies are plotted, the frequency of topologies indicated in the legend are ordered (top to bottom) 
with the exception of the label `other topologies'. The top panel shows the seven most frequently 
occurring topologies originating from stop/sbottom decays while the lower panel represents those 
originating from electroweakino decays. 
\label{fig:missingtopos_stop}}
\end{center}
\end{figure}

Figure~\ref{fig:stop_smodels_summary} demonstrates the existence of a large number of points not 
excluded by \smodels. In order to understand the characteristics of these points and to suggest 
further ways to constrain the non-excluded regions of parameter space, it is interesting to ask 
which SMS topologies not covered by current SMS searches in the \smodels database prevail in 
these points. These are dubbed `missing topologies' and will be discussed in this section in 
details.
The missing topologies are derived purely on the basis of cross section times branching ratio 
(\sigmaXBF) computations and hence do not take into account the sensitivity of the experimental 
searches. For example, in the results that follow, the hadronic decays of W will show up most of 
the times because the branching ratio for W decays to quarks is higher than into leptons, 
however, it is more difficult to beat the backgrounds while searching for hadronic decays.  
Note that it is possible to constrain some of the missing topologies by means of reinterpreting 
the existing experimental searches, thus missing topologies are not always associated with a new 
signature.

The procedure used to derive missing topologies in \smodels is as follows. Each SUSY point leads 
to more than one missing topology.  \smodels sums over the \sigmaXBF for all the topologies irrespective 
of the mass vector of the SUSY particles. Up to 10 such topologies with the highest \sigmaXBF 
are recorded in the output. In the following results, we suppress pair production of LSP which 
occurs as a missing topology. Moreover, we do not consider any initial/final state radiation 
effects.

In Fig.~\ref{fig:missingtopos_stop}, for each non-excluded point, the SMS topology leading to 
the highest \sigmaXBF is chosen. We plot the seven most frequently occurring topologies 
originating from decays of stop/sbottom (upper panel) or electroweakinos (lower panel).
The labels of the plot are written in a simplified notation with respect to the one  used in 
 \smodels ~\cite{Kraml:2013mwa,Kraml:2014sna}, however it is easy to map the current notation 
 to the original one. In this work, every branch is enclosed in parenthesis $()$ and every vertex is 
 separated with a comma ($,$). In both notations, it is assumed that every branch is accompanied 
 with MET at the end of the cascade. The possible origins of these topologies are explained in 
Table~\ref{tab:smodels-missing-illustration}. Note that with this notation we do not distinguish 
particles and antiparticles and a sum over light quarks and light leptons is understood. In 
principle, decays apart from those illustrated in the table can contribute to the missing 
topologies however, because \smodels does not keep track of the SUSY particles but only of their 
masses, this information is lost in the process of decomposition. Let us first concentrate on 
the upper panel of Fig.~\ref{fig:missingtopos_stop}. The most frequently occurring topology is 
{\tt (b),(t)}, meaning b + t + MET. The topology occurs via asymmetric decays of pair 
produced stop (sbottom) with one of the stops (sbottoms) decaying directly to top (bottom) + LSP 
while another stop (sbottom) decays to bottom (top) via chargino. The decay of chargino itself 
is invisible when the chargino - LSP mass gap is less than 5 GeV. Given the extreme degeneracy 
of the chargino - LSP masses the frequency of this topology is hardly a surprise. In general, 
this topology is difficult to be constrained by the current LHC searches as it leads to 2 b jets 
+ one lepton + MET or 2 b jets + 2 jets + MET final state. This is a topology with low jet 
multiplicity while most of the current searches for stops (sbottoms) require high multiplicity 
of light jets. Recently, limits for this mixed topology were made available by both the 
ATLAS~\cite{Aad:2014bva} and CMS~\cite{Khachatryan:2015pwa} collaborations. These limits assume 
exactly 50\% branching ratios for each decay mode. The limits thus obtained are  a sum over 
symmetric and asymmetric decay modes and do not represent upper limits on b + t + MET final state 
alone. For this reason, these limits are not applicable to our topology.  

Other topologies involve the cascade decay of one or both pair-produced squarks into a quark and 
a heavier neutralino (chargino) which then decays into a LSP and an off-shell $Z$ ($W$). It is 
important to notice that the topologies {\tt (t)(t, $\Whad/\Zhad$)}, {\tt 
(t, $\Whad/\Zhad$)(t, $\Whad/\Zhad$)}, {\tt (b, $\Whad/\Zhad$)(t)} for which stops 
contribute, are suppressed for $\mstl > 1$ TeV. This is simply because the production cross 
section of stops heavier than 1 TeV is extremely small. Thus the only relevant topologies are 
those resulting from sbottom decay, it occurs only when the right handed sbottom is lighter than 
the stops. Note that when $W^*_{had}$ or $Z^*_{had}$ are found in a  missing topology, there is also  the same topology with  $W^*_{lep}$ or  $Z^*_{lep}$ with a cross section reduced by the relative leptonic to hadronic branching ratio of the gauge boson. Despite the smaller cross section the  leptonic final states typically have a much better signal to background ratio.
 
The most frequent topology resulting from electroweakino production and decays represented in 
the lower panel of Fig.~\ref{fig:missingtopos_stop} corresponds to associated production of the 
LSP with a chargino (heavier neutralino) decaying via an off-shell $W$ ($Z$). Other topologies 
get their dominant contributions from the decays of at least one heavier neutralino or chargino, 
as detailed in Table~\ref{tab:smodels-missing-illustration}. The very mixed nature of these 
electroweakino lead  to a sizeable production cross sections despite heavier masses.
\clearpage
\begin{table}[h!]
\begin{center}
\tabulinesep=1.4mm
\hspace*{-7mm}
\begin{tabu}{|l|l|} 
\hline
Topology                                                     &   Decay        \\
\hline
(b)(b, $\Whad/\Zhad$) &   $\sbottom1\,\sbottom1 \to b \neutl\, b\, \neuth \to b \neutl\, b\, \Zhad \neutl $\\
\hline
 \multirow{2}{*}{(b)(t)} &  $\stopl\,\stopl \to t \neutl\, b\, W_{soft}\, \neutl $                \\
 & $\sbottom1\,\sbottom1 \to b \neutl\, t\, W_{soft}\, \neutl $                \\
\hline
(b, $\Whad/\Zhad$)(b, $\Whad/\Zhad$) & $\stopl\,\stopl \to b\, \Whad\, \neutl\, b\, \Whad\, \neutl $                  \\ 
\hline
 \multirow{2}{*}{(t)(t, $\Whad/\Zhad$)} &   $\stopl\stopl \to t \neutl\, t \neuth \to t \neutl\, t \Zhad \neutl $\\
 & $\stopl\stopl \to t \neuth\, t \neuth \to t \Zhad \neutl\, t \Zhad \neutl$ \\
 \hline
 \multirow{2}{*}{(b)(t, $\Whad/\Zhad$)} & $\stopl\stopl\to t\neutl\, t\neuth\to t \neutl\, t \Zhad \neutl$   \\
 & $\stopl\stopl\to t \neuth\, t \neuth \to t Z^*_{\nu \nu} \neutl\, t \Zhad \neutl$   \\
\hline
 \multirow{2}{*}{(t, $\Whad/\Zhad$)(t, $\Whad/\Zhad$)} & $\stopl\,\stopl \to t \neuth\, t \neuth \to t \Zhad \neutl\, t \Zhad \neutl$ \\
& $\sbottom1\,\sbottom1 \to t \charginol\, t \charginol \to \, t \Whad\, \neutl\, t \Whad\, \neutl $ \\
\hline
 \multirow{2}{*}{(b, $\Whad/\Zhad$)(t)} &  $\sbottom1\sbottom1 \to b \neuth\, t \charginol \to b \Zhad \neutl\, t W_{soft} \neutl$ \\
 &  $\stopl\stopl \to b \charginol\, t \neutl \to b W_{had} \neutl\,t  \neutl$ \\
\hline
(b)(b, $\Zlep$) &  $\sbottom1 \sbottom1 \to b \neutl\, b \neuth \to b \neutl\, b \Zlep \neutl $ \\
\hline
 \multirow{2}{*}{(b)(b, $\gamma$)} &   $\sbottom1\,\sbottom1 \to b \neutl\, b\, \neuth \to b \neutl\, b\, \gamma \neutl $ \\
 &  $\sbottom1\,\sbottom1 \to b \neuth\, b\, \neuth \to b Z^*_{\nu \nu} \neutl\, b\, \gamma \neutl $ \\
\hline
\hline
(inv)($\Whad/\Zhad$) &      $\charginol\,\neutl \to \Whad\, \neutl\, \neutl$            \\
& $\widetilde{\chi}^0_j\,\neutl \to \Zhad\, \neutl\, \neutl$            \\
\hline
(W)(W) &  $\widetilde{\chi}^{\pm}_2\,\widetilde{\chi}^0_j\, \to W\, \neutl\, W\, \charginol \to W, \neutl\, W\, W_{soft}\, \neutl$                             \\
\hline
(inv)(photon) &    $\charginol\,\neuth \to W_{soft}\, \neutl \,\gamma\, \neutl$              \\
\hline
(inv)(W) &     $\charginol\,\widetilde{\chi}^0_3\, \to W_{soft}\, \neutl\, W\, \charginol \to W_{soft}\, \neutl\, W\, W_{soft}\, \neutl$             \\
\hline
(inv)($\Wlep$) &      $\charginol\,\neutl \to \Wlep\, \neutl\, \neutl$  \\
\hline
(inv)($\Zlep$) &        $\charginol\,\neuth \to W_{soft}\, \neutl\, \Zlep \neutl$            \\
\hline
 \multirow{2}{*}{($\Whad/\Zhad$)($\Wlep$)} & $\neuth\,\charginol \to \Zhad\, \neutl\, \Wlep \neutl$            \\
 & $\charginol \charginol \to \Whad \neutl\, \Wlep \neutl$ \\
\hline
(inv)(b, b) &       $\neuth\,\charginol \to Z^*_{bb}\, \neutl\, W_{soft} \neutl$            \\
\hline

\end{tabu}
\caption{Missing topologies represented in Figs.~\ref{fig:missingtopos_stop},\ref{fig:missingtopos_sbot_freq}
and \ref{fig:missingtopos_neut_freq}, written in \smodels notation and the corresponding physical process. 
$W_{soft}$ represents soft decays of W, which are undetected. $\Zlep$ and $\Wlep$ represent the sum over 
all three generations of leptons, $\Zhad$ and $\Whad$ represent the sum over 
first two generation quarks.\label{tab:smodels-missing-illustration}}
\end{center}
\end{table}
\clearpage
The main topologies, mono -- W ({\tt (inv)(W)}), mono -- photon ({\tt (inv)($\gamma$)}) and diboson 
({\tt (W)(W)}) \footnote{In principle, SMS results for this topology are available by 
ATLAS~\cite{Aad:2014vma}. However, this result is not yet included in \smodels. The reach of 
this analysis at the moment is very limited, excluding chargino mass up to 180 GeV and reaching 
up to maximum neutralino mass of 30 GeV. Inclusion of this search does not exclude any points as 
the LSP is always heavier than $\sim 40$ GeV for our scenario.} cover different mass range s
for the LSP and this independently of the mass of the stop.
Mono -- W topologies could be used to probe the region $100 < \mlsp < 200$~GeV while the mono--photon 
dominates for higher LSP masses.  Note that the mono -- photon topology is here associated 
with $\charginol+\neuth$ production where the chargino decays invisibly and the 
heavier neutralino decays via a loop-induced decay into the LSP and a photon. Because of the 
small difference between the $\neuth - \neutl$ masses, the loop-induced decay 
can have a large enough branching ratio (typically $\mathcal{O}(5\%)$) to give a signature while the 
products of the three-body decay are too soft to be detected and are thus registered as a pure 
missing energy signature. Note that these photons could have a large $p_T$. The diboson topology 
which arises from heavier chargino decays cover the full mass range, in particular the bino-like 
neutralino branches (with $\mlsp \approx 60$ GeV) where there are no topologies occurring from 
the decays of stops/sbottoms.  Updates of searches like~\cite{Aad:2014vma} in the diboson final 
state from Run-II will thus be important. In fact, reinterpreting some of the existing 
mono -- lepton~\cite{ATLAS:2014wra,Khachatryan:2014tva} and mono -- photon~\cite{Aad:2014tda} 
searches could be useful in further constraining the mono -- W and mono -- photon final states. This 
point is left for further investigation. Such channels could thus indirectly constrain scenarios 
with light stop masses beyond the current reaches of direct stop searches. Note that for 
all topologies involving virtual W's or Z's, it is the hadronic decays of the gauge bosons that 
dominate. However, the leptonic decays will also be present with smaller \sigmaXBF but can in 
principle have higher sensitivity than the hadronic channels.

\begin{figure}[h!]
\begin{center}
\begin{tabular}{ccc}
\hspace{-12mm}\includegraphics[scale=0.5]{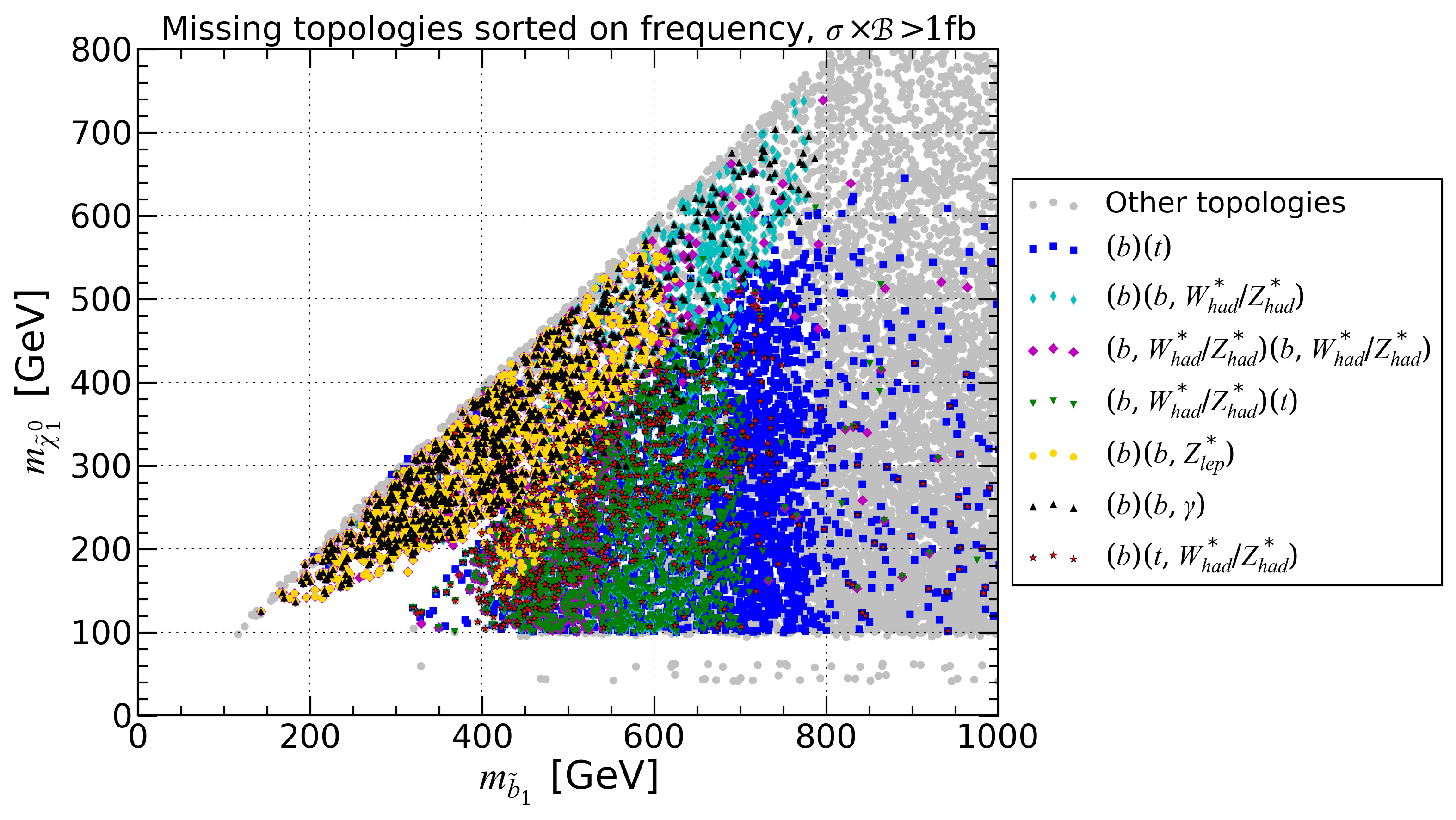}
\end{tabular}
\caption[]{The figure shows missing topologies with cross section greater than 1 fb originating 
from stop or sbottom decays and sorted on their frequency of occurrence.  The seven most frequently 
occurring topologies are illustrated. Topologies originating from electroweakino decays are 
suppressed. 
\label{fig:missingtopos_sbot_freq}}
\end{center}
\end{figure}

\begin{figure}[h!]
\begin{center}
\begin{tabular}{ccc}
\hspace{-12mm}\includegraphics[scale=0.5]{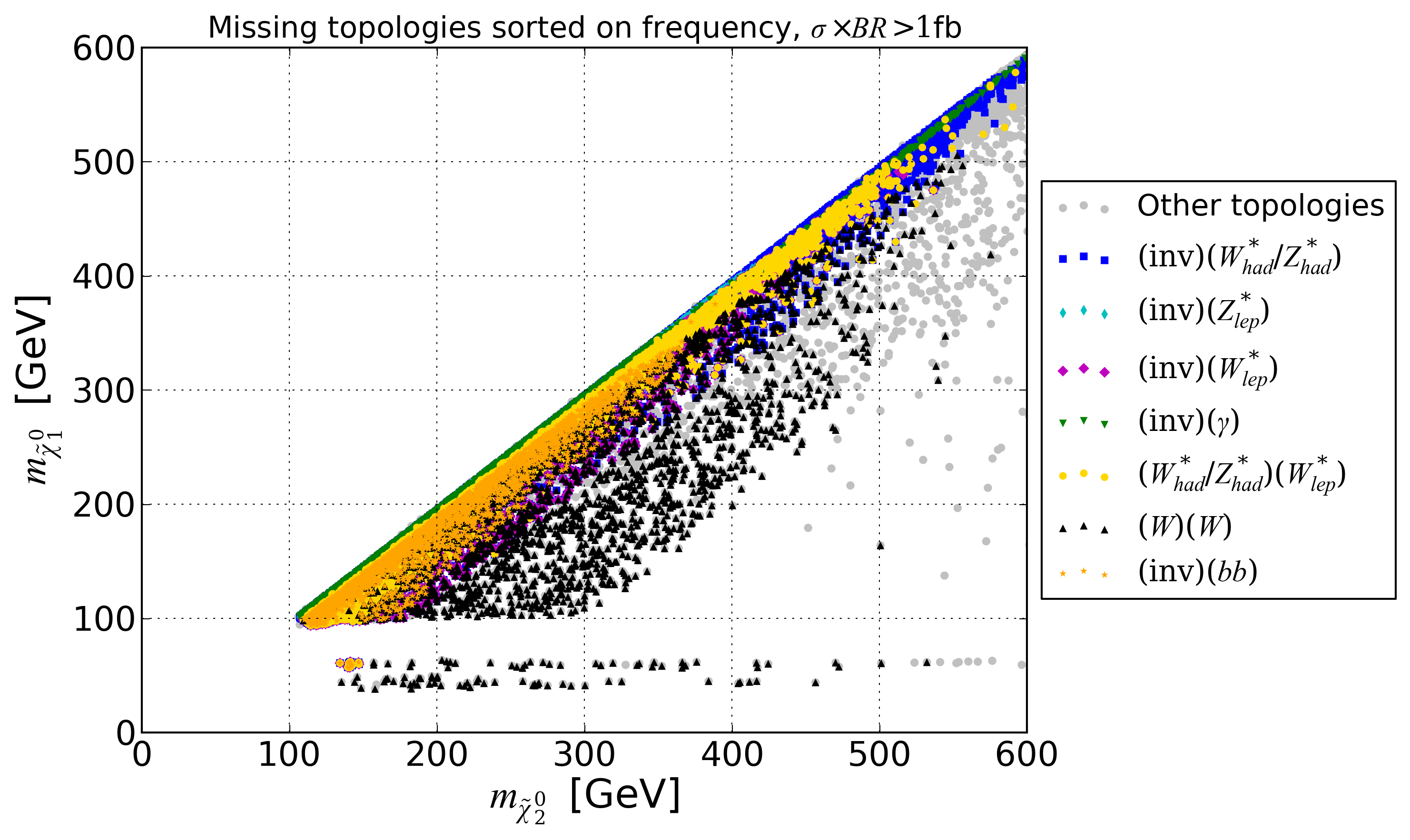}
\end{tabular}
\caption[]{The figure shows missing topologies with cross section greater than 1 fb originating due 
to electroweakino decays sorted on their frequency of occurrence. The minimum cross section of any 
topology plotted is 1fb. The seven most frequently occurring topologies are illustrated. Topologies 
originating from stop/sbottom decays are suppressed.
\label{fig:missingtopos_neut_freq}}
\end{center}
\end{figure}

As explained at the beginning of  this section, each SUSY point leads to more than one 
missing SMS topology, a list of up to 10 such topologies is available from \smodels. In 
Fig.~\ref{fig:missingtopos_stop}, the topologies with the highest cross sections are described. 
The question of which other topologies occur in this scenario, and whether they lead to some 
more sensitive final states still needs to be answered. In order to illustrate this, all topologies 
with a $\sigmaXBF > 1\textrm{fb}$ are sorted 
according to their frequency of occurrence and the seven most frequent topologies are shown in 
Figs.~\ref{fig:missingtopos_sbot_freq} and Fig.~\ref{fig:missingtopos_neut_freq}. Once again, in 
order to make it easy to understand the origin and correlation of the topologies, only 
topologies originating from stop/sbottom decays are plotted in Fig.~\ref{fig:missingtopos_sbot_freq} 
in the $\msbl - \mlsp$ plane, while those from the decays of 
electroweakinos are plotted in Fig.~\ref{fig:missingtopos_neut_freq}. Comparing Fig. 
~\ref{fig:missingtopos_sbot_freq} to the top panel of Fig.~\ref{fig:missingtopos_stop} it is 
clear that along with the dominant topologies, {\tt (b)(t)} and those involving hadronic 
W*/Z* decays that are found in both figures, it is also possible to find other topologies which 
might have a better signal to background ratio e.g. {\tt (b)(b, $\Zlep$)} or {\tt 
(b)(b,$\gamma$)}. The latter again occurs from the loop-induced decay of the second 
neutralino produced in sbottom decay as explained in 
table~\ref{tab:smodels-missing-illustration}. This topology with 2 b jets, 1 photon + MET in the 
final state can be used to constrain the kinematic edge with $\msbl - \mlsp < 50$ GeV.

Missing topologies in Fig.~\ref{fig:missingtopos_sbot_freq} show a strong correlation with the 
sbottom mass. Topologies involving two b jets, MET and a virtual gauge boson 
originating from the decay of sbottoms into a heavier neutralino/chargino occur mostly in the 
region with a small difference between the sbottom and the LSP. Only a few topologies involving 
stop pair production extend beyond $\msbl > 800$ GeV where the sbottom pair production 
becomes too small.

In Fig.~\ref{fig:missingtopos_sbot_freq}, it is difficult to highlight topologies originating from the decays of stops. 
The reason is the large mixing angle in the stop sector due to the Higgs mass constraints. 
In this case, the stop has no preferred decay channel, thus, missing topologies with a large 
$\sigmaXBF$ which naturally result from a large branching ratio in a single channel are absent.

As the $\mc1$ and $\mlsp$ are extremely mass degenerate, we plot the missing topologies in the 
electroweak sector in the $\neuth - \neutl$ plane. Fig.~\ref{fig:missingtopos_neut_freq} shows 
that most of the topologies in this sector are in fact off-shell decays of $W$ and $Z$ bosons 
associated with MET. Clearly, in most of the region where hadronic decays of W or Z 
are dominant, the leptonic decays are also present. Moreover the importance of the diboson 
missing topology is once again evident.  Expectedly, in the region with $m_{\neuth}, m_{\neutl} \gtrsim 400 {\rm GeV}$, 
the leptonic decays become irrelevant due to low \sigmaXBF.

\subsection{Stop NLSP at LHC 14 TeV}

In this section we focus on the region of parameter space where the lightest top squark is the 
NLSP and has a small mass difference with the LSP hence the decay $\stopl \to t 
\neutl$ is not open. If the mas gap $\mstl - \mlsp \lesssim 85$ GeV (which is true 
for most of the allowed points with a stop NLSP) then the 3-body decay $\stopl \to b \, 
W^+ \, \neutl$ is kinematically forbidden and the only allowed decay modes are the 
4-body decay $\stopl \to b \, \neutl \, f \, f^\prime$ and the flavour 
violating decay $\stopl \to c/u \, \neutl$. Because of the small mass gap 
between the light stop and the LSP, the decay products are very soft on average making this 
scenario extremely challenging for the LHC searches. This explains the very low lower bound on 
the light stop mass ($\mstl \gtrsim 275$ GeV) obtained with the LHC 8 TeV data for this specific 
scenario. Hence, it is interesting to explore the prospect of the 14 TeV LHC for this scenario. 
It was shown that using monojet + large missing transverse energy final state a stop mass 
$\gtrsim 300$ GeV will not be ruled out even with 100 fb$^{-1}$ integrated luminosity at 14 TeV 
LHC~\cite{Drees:2012dd}. The same question was also investigated in 
Ref.~\cite{Belanger:2013oka} using $\alpha_T$ and $M_{T2}$. Assuming $\br(\stopl \to c 
\, \neutl)$ = 1 it was shown that the exclusion limit can be extended to a maximum 
of $\sim 450$ GeV (depending on the mass gap between $\stopl$ and 
$\neutl$) with 100 fb$^{-1}$ data.  
Moreover, the use of charm-tagging could prove very useful in the future~\cite{Belanger:2013oka}. 
\begin{figure}[h!]
\begin{center}
\begin{tabular}{cc}
\includegraphics[scale=0.5]{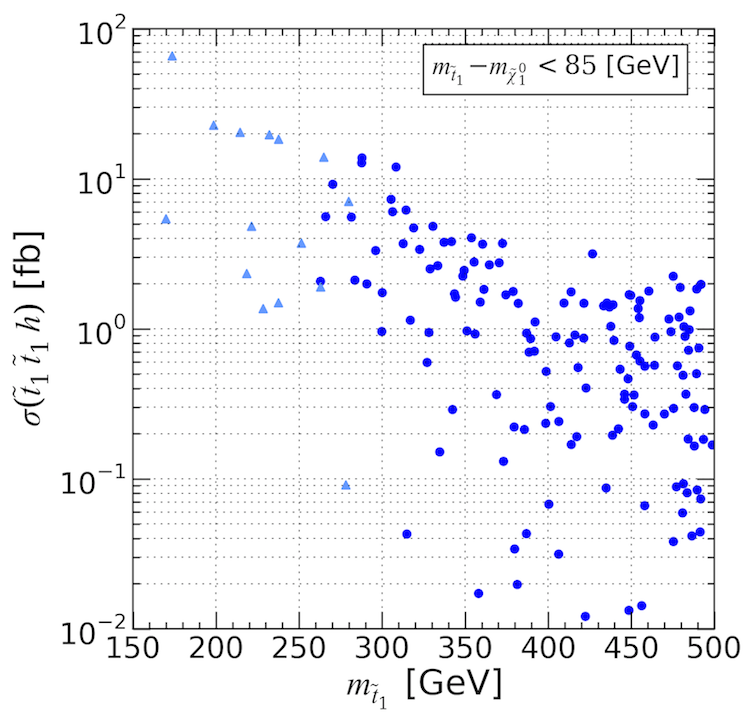} & 
\includegraphics[scale=0.5]{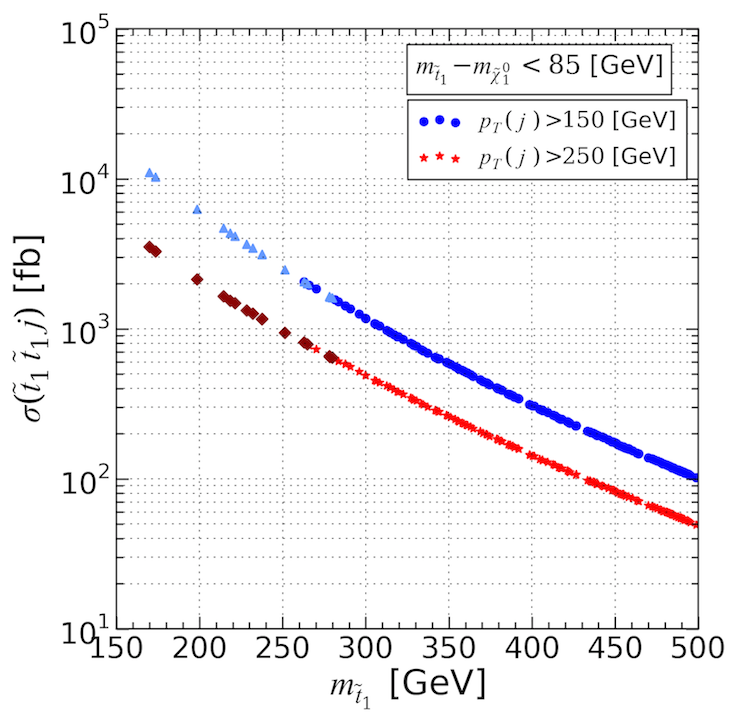}
\end{tabular}
\caption[]{Leading order cross sections for $\stopl \stopl j$ and $\stopl \stopl h$ at the 
14 TeV LHC. MadGraph5\_aMC@NLO version 2.2.2 \cite{Alwall:2014hca} was used for the computation.
The triangles represent  points that are potentially excluded from ATLAS searches for 
$\stopl \to c + \neutl$ and stop four body 
decays.
\label{fig:res-4}}
\end{center}
\end{figure}

As there are not many handles to suppress the SM backgrounds in the stop NLSP scenario, the 
radiation of a light Higgs from one of the stops in stop pair production could provide an additional 
handle to discriminate the signal. It was shown in~\cite{Djouadi:1999dg,Belanger:1999pv} that 
the associated production of a stop pair with a light Higgs could be large in some region of parameter 
space, basically due to a large $\stopl \stopl h$ coupling. However, when taking into 
consideration the mass of the Higgs which requires a contribution from the stops and maximal 
mixing, the $\stopl \stopl h$ vertex is suppressed, hence the $\stopl \stopl h$ 
cross section is expected to be quite low. Moreover this process depends on the parameters of 
the model whereas the cross section for $\stopl \stopl j$ production is expected to be much 
larger and depends only on $\mstl$. In Fig.~\ref{fig:res-4} we show $\stopl \stopl h$ (left panel) 
and $\stopl \stopl j$ (right panel) production cross sections at the 14 TeV LHC for all the 
allowed points with the light stop as the NLSP. The triangles represent potentially excluded 
points due to constraints coming from ATLAS $\stopl \to c + \neutl$ and stop four body 
decays. The production cross section for $\stopl \stopl h$ final state is at least an order of 
magnitude smaller than that for $\stopl \stopl j$. This makes the $\stopl \stopl h$ channel much 
less promising than the $\stopl \stopl j$ final state. Although there can be considerable gain 
in the background reduction if the Higgs in the final state is tagged but the existence of 
irreducible background like $p p \to h Z (\to \nu \nu)$ (with a 14 TeV cross section $\sim 150$ 
fb) makes it extremely challenging. For definiteness, in Table-\ref{benchmarks} we provide a few 
example benchmarks with different $\stopl$ decay modes. We avoid a choice of benchmark point 
within the excluded regions of the existing ATLAS $\stopl \to c \, \neutl$ or stop 
four body decays.

\begin{table}[h!]
\begin{center}
\tabulinesep=1.4mm
\hspace*{-7mm}
\begin{tabu}{|c|c|c|c|c|c|c|} 
\hline
$\mstl$ & $\mlsp$ & $\mc1$ & \br($\stopl \to c \neutl$) & 
\br($\stopl \to b f f^\prime \neutl$) 
& \br($\stopl \to b \charginol$) 
& $\sigma(\stopl \stopl h) (\rm fb)$\\
\hline 
305.3  & 295.3 &  820.7  & 1 & 0  & 0  & 7.3  \\
\hline
372.4  & 364.0 & 1097.0  & 1 & 0  & 0  & 3.7  \\
\hline
328.8  & 301.2 & 1033.6  & 0.41 & 0.59  & 0  & 2.5  \\
\hline
314.3  & 305.1 &  309.5  & 0 & 1  & 0  & 6.2  \\
\hline
308.3  & 260.1 &  264.0  & 0 & 0  & 1  & 12.0 \\
\hline
353.7  & 319.8 &  322.8  & 0 & 0  & 1  & 4.1  \\
\hline
\end{tabu}
\caption{A few benchmark points to show specific examples of different stop decay modes and 
also the 14 TeV $\stopl \stopl h$ production cross section in fb. While for 
the first three benchmarks the lightest stop is the NLSP, for the final three points the 
lightest chargino is the NLSP.
\label{benchmarks}}
\end{center}
\end{table}

In the first two benchmarks 
$\stopl$ exclusively decays to $c \neutl$ and in the following two points 
$\stopl$ exclusively decays to the 4-body final state. In the final two benchmarks $\stopl$ is 
not the NLSP - the $\charginol$ and $\neuth$ lie beneath $\stopl$ in the spectrum. 
As a consequence, the $\stopl$ 
exclusively decays to $b \charginol$. The chargino in this case will eventually decay to 
$f f^\prime \neutl$ giving rise to again a 4-body decay of $\stopl$ but because of the 
existence of an on-shell chargino in the decay chain, the kinematics will be different. 
In summary, the stop pair production in association with an extra hard jet will be the 
most promising channel for probing the stop NLSP region with small stop-neutralino mass gap. 
The prospect of stop pair production process associated with a Higgs boson does not look 
encouraging mainly because of the small production cross section. We find that with the Higgs mass constraint, 
the stop composition is such that the $\stopl\stopl h$ coupling is small. This means that the  
$\stopl\stopl h$ cross-section is suppressed even for light stops, the suppression being dynamic and 
not so much due to kinematics. The existence 
of a Higgs in the final state may provide an additional handle to combat backgrounds, 
hence a more focused study may be worthwhile. 
As far as the decay of stop is concerned, there are three distinct categories where the final state objects 
and/or the kinematics are different. 
Hence, dedicated searches at the 14 TeV LHC for each of them should be carried out.
\subsection{Decays of heavier stop}
\label{sec:heavystop}

\begin{figure}[h!]
\begin{center}
\begin{tabular}{cc}
\includegraphics[scale=0.45]{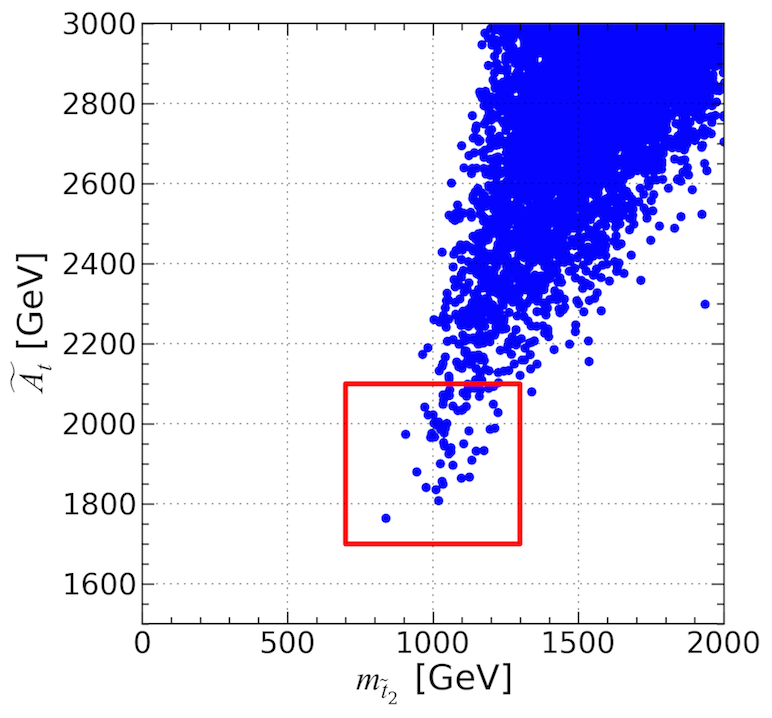}&
\includegraphics[scale=0.45]{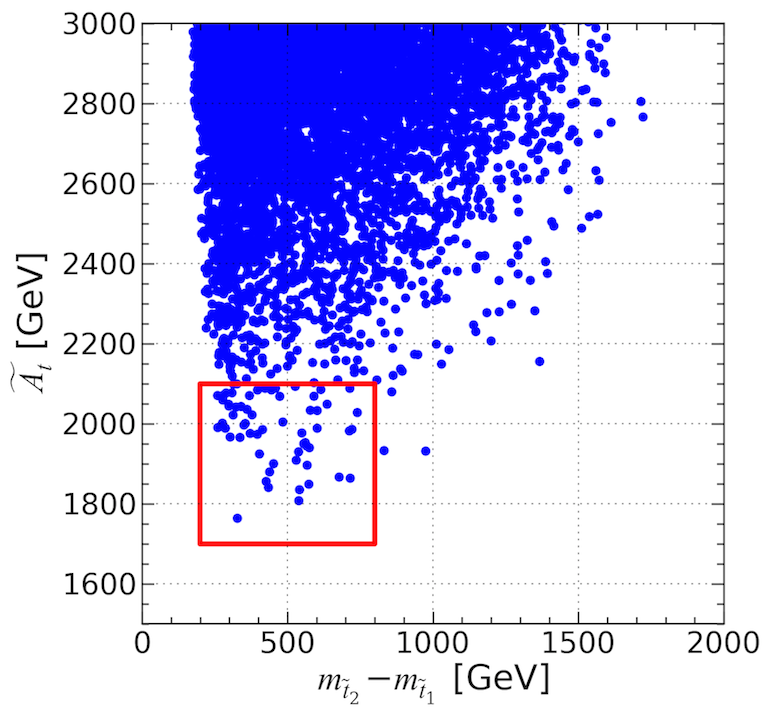}
\end{tabular}
\caption[]{ The allowed pMSSM points in the 
$\tilde{A}_t - \msth$ (left panel) and 
$\tilde{A}_t - $ ($\msth$ - $\mstl$)  
(right panel) planes. 
\label{fig:res-5}}
\end{center}
\end{figure}

Searches for the heavier stop could provide an alternative for probing the light stop scenarios, 
the main issue for exploiting these searches at the LHC remains the mass scale of the heavier 
stop. Due to the large radiative corrections required to achieve the correct Higgs mass in the 
MSSM either large $A_t$ or heavy stop masses is required. A large value of $A_t$ introduces a 
large splitting in stop masses and pushes the heavier stop above the TeV scale.
In the left panel of Fig.~\ref{fig:res-5} we plot the allowed points in the $\tilde{A}_t -
\msth$ plane, while in the right panel the same set of points are plotted against the mass 
difference between two stops ($\msth-\mstl$). Recall that $\tilde{A}_t$ is related to the 
naturalness of the SUSY spectrum (see Eq.~\ref{stop-naturalness}). Fig.~\ref{fig:res-5} shows 
that the value of $\tilde{A}_t $ is constrained to be approximately above 1.8 TeV. This, in 
turn, gives a lower bound on the fine-tuning parameter $\Delta \gtrsim 50$ (assuming 
$\Lambda_{\rm mess}$ = 20 TeV) which amounts to $\sim 2\%$ tuning.

An optimistic scenario with $\Delta \sim 50$ leads to $\msth \sim 1$ TeV and $(\msth 
-\mstl) \sim 500$ GeV (the region enclosed by the thick red line). Thus, an interesting outcome 
of our analysis is the possibility of a SUSY spectrum with a heavier stop mass around a TeV with 
a large mass gap ($\sim 500$ GeV) with the lighter stop. The discovery potential of this 
scenario at the 14 TeV LHC will in general depend on the decay channel of $\stoph$ 
which in turn depends on the masses of the other SUSY particles. 
We will consider two specific decay modes $\stoph \to 
\stopl Z$ and $\stoph \to \stopl h$ which are particularly interesting and have in 
general large branching ratios (as we will show below). The couplings 
$\stoph$--$\stopl$--$Z$ (in the limit when the mass splitting between 
$\stoph$ and $\stopl$ is large and $M_A, M_H \gg M_Z$) and 
$\stoph$--$\stopl$--$h$ can be written as \cite{Djouadi:1999dg,Belanger:1999pv},
\bea
\label{t2t1z}
\lambda_{\stoph \stopl Z}  &\approx& \dfrac{g}{2 M_W} \, m_t X_t \, \, , \\
\lambda_{\stoph \stopl h}  &\approx& 
2 (\sqrt{2} G_F)^{\frac{1}{2}} M_Z^2  \, \times \nonumber \\[2mm]
&& \hspace{-2cm}\bigg[
(\dfrac{2}{3} \sin^2\theta_W - \dfrac{1}{4}) \cos (2\beta)  \, \sin (2 \theta_t) 
+  \dfrac{1}{2} \dfrac{m_t}{M_Z^2} \cos (2\theta_t) \, X_t \bigg] 
\eea
The mixing angle between the left and right handed top squarks, $\theta_t$ and the 
mixing parameter $X_t$ are defined through the mass matrices~\cite{Djouadi:1999dg},
\bea
M_{\widetilde{t}}^2 &=& 
\left( \begin{array}{ccc}
m_{LL}^2 + m_t^2 & m_t X_t \\
m_t X_t & m_{RR}^2 + m_t^2 \\
\end{array} \right) \nonumber \\
X_t &=& A_t - \dfrac{\mu}{\tan\beta} \nonumber \\
m_{LL}^2 &=& m_{\widetilde{Q}_3}^2  + 
(\dfrac{1}{2} - \dfrac{2}{3} \sin^2 \theta_W)\cos 2\beta \, M_Z^2 \nonumber \\
m_{RR}^2 &=& m_{\widetilde{U}_3}^2  + 
\dfrac{2}{3} \sin^2 \theta_W \cos 2\beta \, M_Z^2 \\
m_{\tilde{t}_{1,2}}^2 &=& m_t^2 + \frac{1}{2} \left[ 
m_{LL}^2 + m_{RR}^2 \mp \sqrt{
(m_{LL}^2 - m_{RR}^2)^2 + 4m_t^2 X_t^2 } \right]    \nonumber \\
\sin 2\theta_t &=& \frac{2 m_t X_t} { \mstl^2
-\msth^2 } \nonumber  \ \ , \ \   \nonumber
\cos 2\theta_t = \frac{m_{LL}^2 -m_{RR}^2} 
{\mstl^2 -\msth^2}  \, \, .
\eea
%

\begin{figure}[t!]
\begin{center}
\begin{tabular}{ccc}
\includegraphics[scale=0.5]{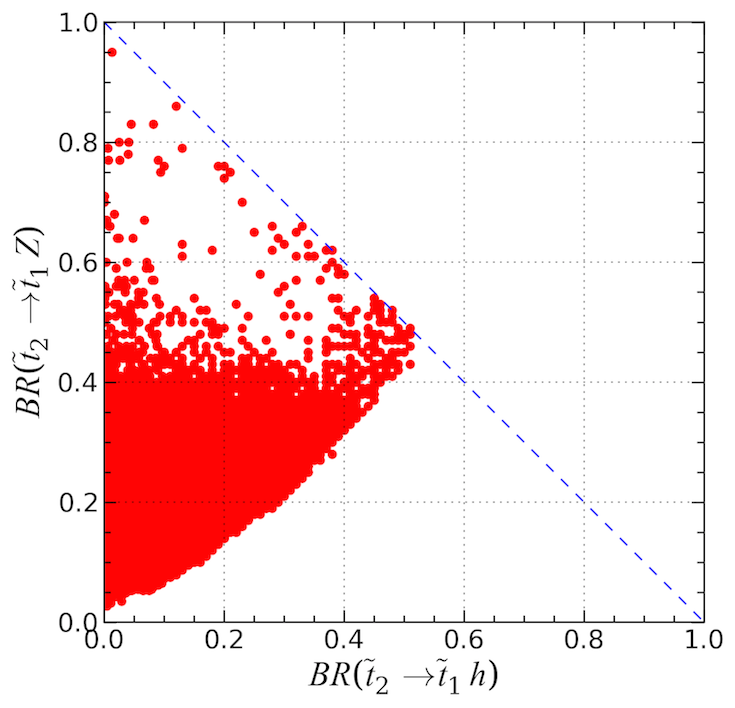}&
\includegraphics[scale=0.5]{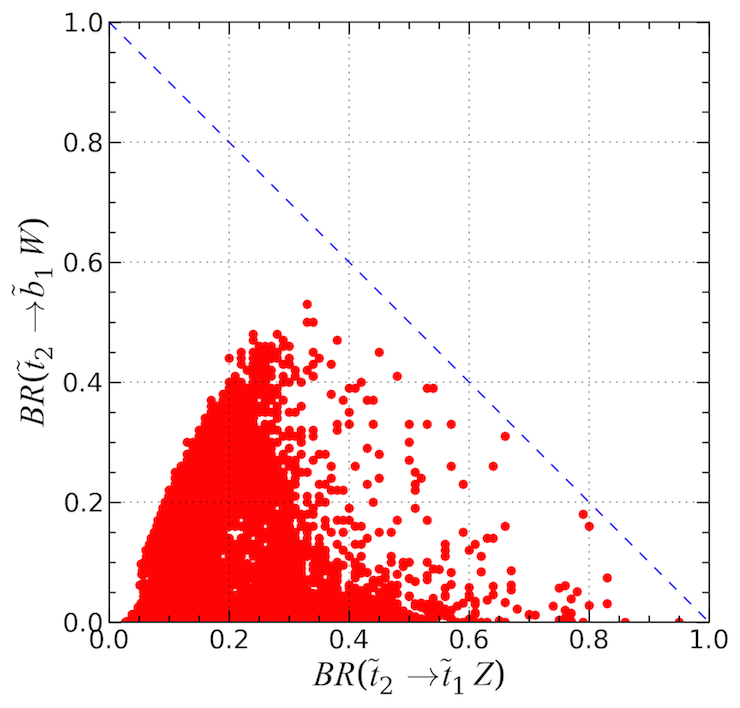}
\end{tabular}
\caption[]{Left panel: branching ratio of $\stoph \to \stopl h$ vs. that of 
$\stoph \to \stopl Z$. 
Right panel: branching ratio of $\stoph \to \stopl Z$ vs. that of 
$\stoph \to \stopl W$. The region only below the blue dashed line is 
physical.
\label{fig:res-6}}
\end{center}
\end{figure}

Eq.~\ref{t2t1z} shows that the $\stoph$--$\stopl$--$Z$ coupling will be large if 
$X_t$ is large (which is required for the Higgs mass unless both stops are extremely heavy). The 
$m_t$ enhancement can also be understood by noting that in the large energy limit (large mass 
splitting between $\stoph$ and $\stopl$ ) the $Z$ boson can be replaced by one 
of the would-be goldstone boson using the goldstone boson equivalence theorem. The 
$\stoph$--$\stopl$--$h$ coupling has slightly more structure and has both the 
F-term and D-term contributions. However, this coupling can also be quite large in some part of 
the parameter space. Hence, these two decay modes are well motivated from the fact that large 
$X_t$ is required for the Higgs mass. Of course, the presence of other particles below the 
$\stoph$ mass makes the picture more complicated and many other decay modes can 
contribute. In Fig.~\ref{fig:res-6} we show the branching ratios into gauge and Higgs boson for 
all the allowed points. One can see that for a large fraction of points the sum of 
$\br(\stoph \to \stopl \, Z)$ and $\br(\stoph \to 
\stopl \, h)$ is quite large supporting our analytic expectation.

Hence, the pair production of $\stoph$ with the subsequent decays $\stoph \to 
\stopl \, Z$ and $\stoph \to \stopl \, h$ can be a very interesting 
channel to look at. The 14 TeV prospect of this scenario was studied in~\cite{Ghosh:2013qga} by 
one of the authors. It was shown that a 4--5$\sigma$ signal can be observed for $\sim$ 1 TeV 
$\stoph$ with 100 fb$^{-1}$ integrated luminosity. We refer the readers to~\cite{Ghosh:2013qga} 
for further details.

\subsection{Decays of heavy Higgs} 

In this section we turn our attention to the neutral heavy Higgs bosons $H$ and $A$.  An 
important result of our analysis is that the mass of the heavy neutral scalar $H$ is constrained  
to be larger than $\sim 450$ GeV. 
Note that this lower bound arises only after imposing constraints from the light Higgs signal 
strengths and Heavy Higgs searches implemented in HiggsSignals and HiggsBounds . Moreover the $A$ 
is quite degenerate with the $H$, the mass difference being almost always less than $5$ GeV, 
which is less than one percent of the common mass. For such heavy Higgses, the widths are $\sim\, 0.05 m_{H}, m_{A}$
from SM fermions alone~\cite{Djouadi:2015jea}. These widths can only increase when decay channels 
into sparticles are included. 
Thus the mass difference  between $H$ and $A$, is always comparable or smaller than their widths.
\comment{Even if we consider only the 
decays of $H$ and $A$ into the SM fermions, the widths are $\sim 0.05 m_{H}, m_{A}$~\cite{Djouadi:2015jea}, 
whereas once the super particle channels open up it can only increase. The major conclusion 
then being that the mass difference between $H$ and $A$, is comparable or smaller than the width 
of the either.} It is well known that the 
agreement of the observed signal strengths with those expected in the SM, actually forces the 
global fits to the alignment region where $|(\beta - \alpha)| \sim \dfrac{\pi}{2}$. As a result 
the gauge boson couplings are severely suppressed for the $H$. Hence, even with the large mass 
of $H$ and the enhancement factor in the $VV$ decay width due to the decays in the longitudinal 
$V$ bosons, the branching ratio into vector bosons is not above one percent. Recall that the 
$AVV$ vertex is absent at tree level. We have also checked that the decays $H \to hh$ 
and $A \to Z h$ have branching ratios smaller than one percent for all our points. Thus, 
the only relevant tree level decay modes can be into standard model fermions and sparticles: the 
sfermions and the electroweakinos. Moreover the branching ratios should be similar for the $H$ 
and the $A$.

\begin{figure}[h!]
\begin{center}
\begin{tabular}{cc}
\includegraphics[scale=0.45]{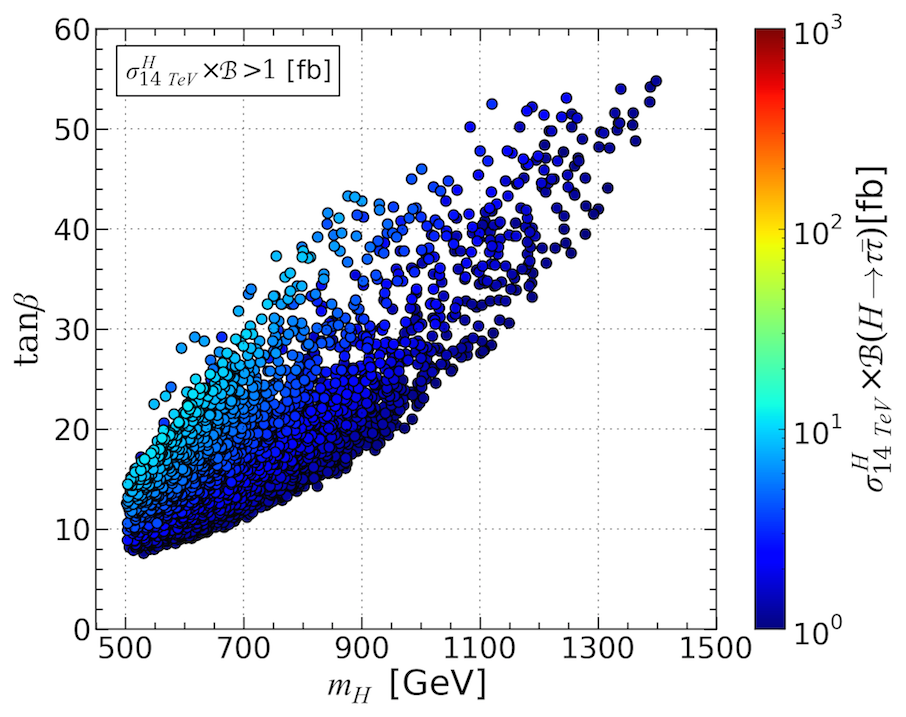}&
\includegraphics[scale=0.45]{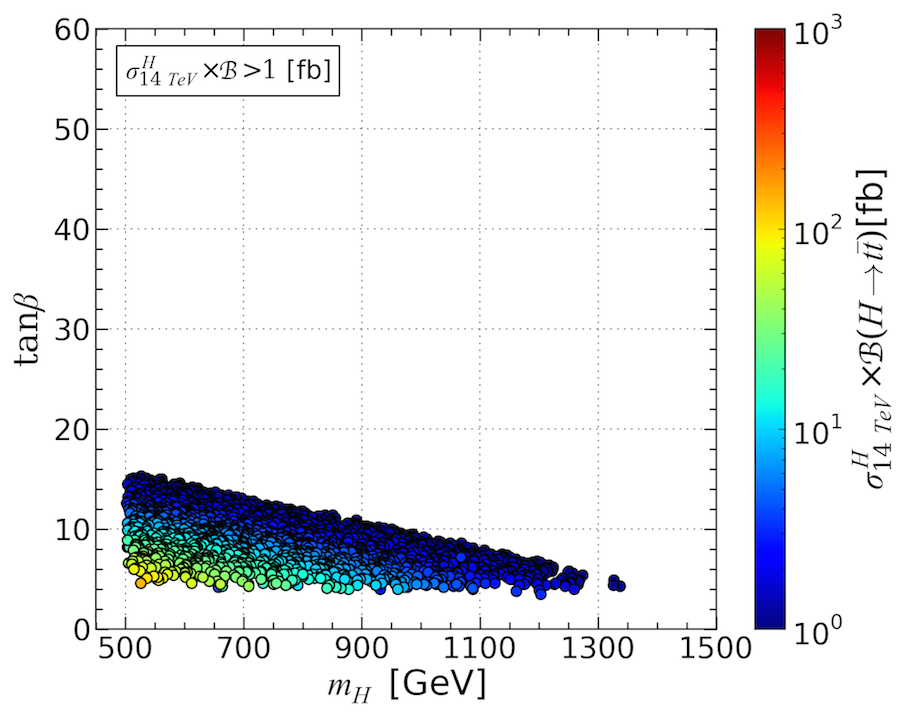}
\end{tabular}
\caption[]{The product of 14 TeV cross section of the CP-even heavy Higgs and its  branching 
ratio to $\tau \tau$ final state (left panel), $t\bar{t}$ final state (right panel) in the 
$\tan\beta - M_A$ plane. 
\label{fig:H_xsec}}
\end{center}
\end{figure}

The large mass of the $H$ means that the decay to the $t \bar t$ final state is now 
kinematically allowed.  As a result it can be the dominant decay mode for small values of 
$\tan\beta$ and the $\br(H \to t \bar t)$ can be as high as $\ge 80 \%$ for $\tan \beta 
\lsim 5$. With increasing values of $\tan\beta$, $\br(H \to t \bar t)$ drops gradually 
and the $H \to b \bar b$ decay mode starts dominating. These two branching ratios sum up 
to about 80\% in large part of the parameter space. The remaining 20\% is mostly taken up by 
$\br(H \to \tau \bar \tau)$ decays. The SM decay modes are however suppressed when decay 
channels into SUSY particles become important as will be discussed below.
In the left panel of Fig.~\ref{fig:H_xsec} we show 
the 14 TeV cross sections for $H$ production ($\sigma^{H}_{14 \rm TeV}$) times the branching 
ratio to the $\tau \, \tau$ and $t\bar{t}$ final states for the allowed points for which 
$\sigma^{H}_{14 \rm TeV} \times \br(H \to f\bar{f})$ is greater than 1 fb. The production 
cross section were computed with SusHi-1.5.0~\cite{Harlander:2002wh,Harlander:2003ai,Aglietti:2004nj,
Harlander:2005rq,Bonciani:2010ms,Degrassi:2010eu,Degrassi:2011vq,Degrassi:2012vt,Harlander:2012pb}. 
The decays of $H/A$ into $\tau$ 
pair clearly offer a search channel with $\sigma^{H}_{14 \rm TeV} \times \br$ that can reach ${\cal O} (10)$ fb, 
for $\tan \beta \lsim 20$ and $m_{H} \sim 1 $ TeV. For the $t \bar t$ final state $\sigma^{H}_{14 \rm TeV} \times \br(H 
\to t \bar t)$ can be as high as 10's of fb and higher for low $\tan \beta \lsim 15$ and 
$m_{H} \lsim 1000$ GeV.

\begin{figure}[h!]
\begin{center}
\begin{tabular}{cc}
\includegraphics[scale=0.45]{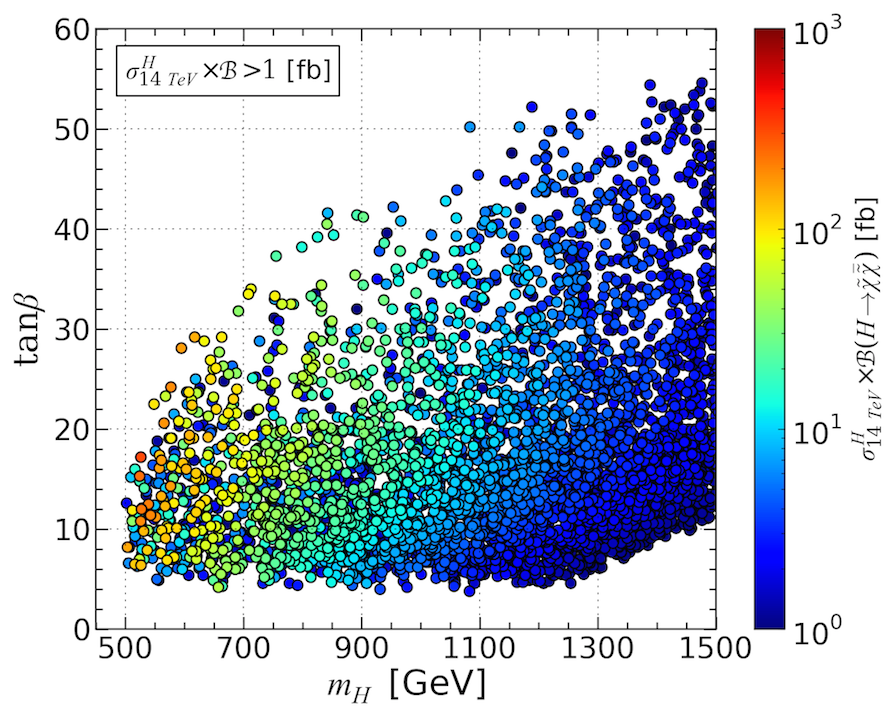}&
\includegraphics[scale=0.45]{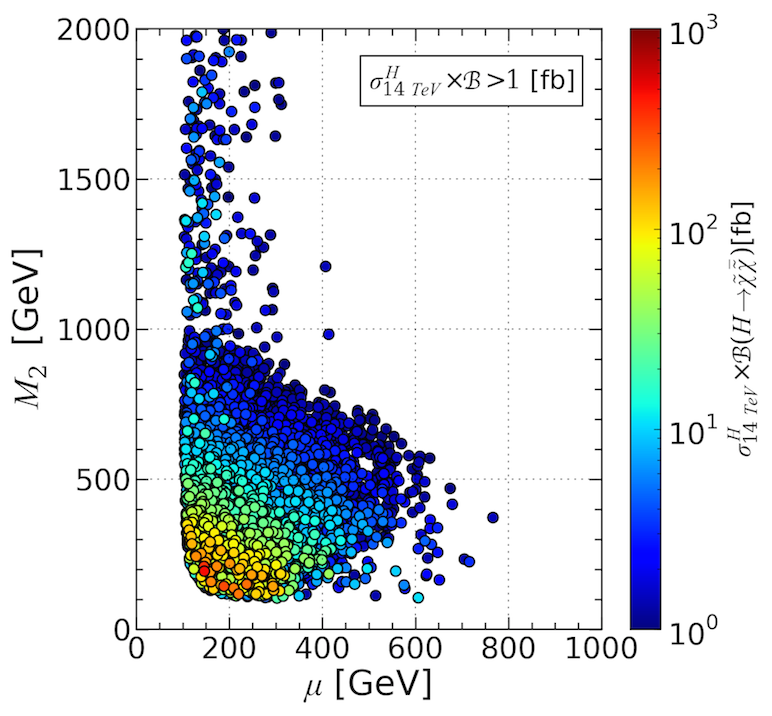}
\end{tabular}
\caption[]{The product of 14 TeV cross section of the CP-even heavy Higgs and its  branching 
ratio to electroweakino final state in the $m_H -- \tan\beta$ plane (left panel) and in the 
$M_2 -- \mu$ plane (right panel). 
\label{fig:H_ewk}}
\end{center}
\end{figure}

The decay of the heavy Higgs $H$ to SUSY final states can also be quite important.  In 
Fig.~\ref{fig:H_ewk} we plot the product of cross section times branching ratio for $H$ to the 
electroweakino final states \footnote{There is a small fraction of $H$ that decays to light 
stops as well. However, this channel is not always kinematically open and most of the times 
$\br(H \to \stopl \stopl^*) < 0.1$.}. All channels where the heavy 
Higgs can decay to charginos or neutralinos are summed over. The `invisible' decays of the $H$ include  $\neutl 
\neutl$  as well as the decay into heavier states when the mass difference 
between the sparticle (eg. $\neuth$) and the LSP is small: $\lsim 5$ GeV. We find that the total 
`invisible' branching ratio of the $H$ is always less than $30 \%$ and mostly less than $10 \%$ 
while the visible decays into electroweakinos can reach 80\%. The branching ratios for the 
$A$ are also similar to those for the $H$ presented in the figure. In fact, apart from the 
production cross section, the phenomenology of $H$ and $A$ for these allowed points are rather 
similar. It is also well known that for the same mass, the gluon fusion production 
cross section can be higher for A than for H,  the exact values depending on the masses of 
the squarks.

There have been dedicated discussions of the Higgs sector of the MSSM, allowed after the LHC 
Run-I, beginning from the analysis of the partial first data set~\cite{Arbey:2013jla} in the 
pMSSM framework to the more recent analyses ~\cite{Djouadi:2015jea,Bhattacherjee:2015sga} in the 
hMSSM and pMSSM scenarios respectively which  
focus on searches for the Heavy Higgses at the LHC 14 TeV.

Obviously, as the above discussion shows the $t \bar t$ final state is perhaps the most crucial 
in the low $\tan \beta$ range and the $\tau \bar\tau$ in the large $\tan \beta$ range. While the 
$\tau \bar \tau$ channel has received a lot of attention in the past, the LHC Run-I results~\cite{Khachatryan:2014wca} have 
forced attention to be focused on the $t \bar t$ final state. It has been known since a long 
time that for this final state, interference with the $t \bar t$ QCD background gives a very 
characteristic peak-dip structure~\cite{Gaemers:1984sj,Dicus:1994bm}. The feasibility of using 
it to isolate the signal from the background as well as the difference in the spin spin 
correlations between the $t$ and $\bar t$ for the background and the resonant signal have been 
discussed in the literature~\cite{Bernreuther:1997gs,Frederix:2007gi,Barcelo:2010bm}. In the CP 
conserving case, the $H$ and $A$ amplitudes do not interfere but still the presence of almost 
degenerate $H,A$ can degrade the effect.  The peak-dip structure seems to be subdominant to the 
effect of higher order corrections~\cite{Moretti:2012mq} and more intricate cuts may have to be 
devised to enhance the resonant Higgs contribution. Moreover a recent analysis \cite{Craig:2015jba} concluded that 
this peak-dip structure will get degraded due to the limited resolution of the $t \bar t$ 
invariant mass, the statement being even more true with the presence of degenerate $H$ and $A$ 
as would be the case here. Hence kinematic cuts which exploit the effect of the spin-spin 
correlations for the $t \bar t$ produced in the $H$ decay would be necessary (see for example 
~\cite{Barger:2006hm}). A simple analysis of ~\cite{Djouadi:2015jea} which includes such cuts, 
shows that at the LHC 14TeV one could be sensitive to $\tan \beta \sim 6$ for $m_{H} \sim 500 $ 
GeV and to $\tan \beta \sim 1$ for $m_{H} \sim 1 $ TeV.  More analyses, to improve the 
sensitivity of this channel are required, see for example~\cite{Craig:2015jba,Hajer:2015gka}. 
 
Of some interest are the decays into the electroweakinos where the product of cross section 
times branching ratios can reach values as high as $10^{2}$ to $10^{3}$ fb for $H$ masses up to 
$700- 1000$ GeV and $\tan \beta \lsim 20-30$. Thus we find that this channel can offer 
interesting search possibility for the $H/A$. Note that the general conclusion of 
~\cite{Djouadi:2015jea} that the parameter ranges which gives large branching into these 
channels have been ruled out by the LHC trilepton constraints does not apply in our analysis 
where $M_{1}, M_{2}$ values are not related. The large values of the cross section times 
branching are concentrated in the low $M_{2}$-$\mu$ region and hence the chargino/neutralinos 
produced are likely to give rise to final states with real or virtual $W/Z$. However, the 
topology for final states resulting from the $H$ decay into electroweakinos, followed by their 
cascade decays, will depend on their couplings and on the masses  which are strongly influenced 
by the relic density upper limit. A detailed investigation of the potential to use electroweakino 
decay channels to extend the reach for heavy Higgs, a topic that was hardly addressed 
~\cite{Bisset:2007mi} is beyond the scope of this paper.

As mentioned already the $H$ has an invisible branching ratio which can be upto $0.3$ but is 
mostly less than $0.1$. One can in fact search `directly' for such an invisible $H$ at the LHC 
via the associated production of the $H$ with a vector boson~\cite{Godbole:2003it} or production 
of $H$ via the vector boson fusion ~\cite{Eboli:2000ze}. Indeed currently bounds exist on the 
invisible branching ratio of the $126$ GeV Higgs by both the CMS~\cite{Chatrchyan:2014tja} and 
ATLAS~\cite{Aad:2014iia,Aad:2015uga}, using both the modes. For a $120$ GeV Higgs boson, the 14 
TeV LHC should be able to probe a value of the branching ratio as low as 
$0.17$~\cite{Ghosh:2012ep}.  Earlier projections~\cite{Godbole:2003it,Eboli:2000ze} had looked 
at larger values of the scalar mass. Unfortunately, these channels will be of not much use in 
the present case for the $H$ as its couplings to a $VV$ pair are highly suppressed. Hence the 
production of $H$ in gluon fusion with associated jets followed by the decay of $H$ in invisible 
channel~\cite{Djouadi:2012zc} or associated production of $H$ with a $t \bar t$ pair followed by 
an invisible decay of the $H$, are the two possibilities for such an invisibly decaying $H$. The 
estimates of the expected rates for the latter channel presented in~\cite{Craig:2015jba} 
assuming an invisible branching ratio of $1$, shows that the search in the channel $t \bar t$ 
pair + MET will be a challenging one. As far as an invisibly decaying $A$ is 
concerned, for the same mass and coupling strengths, the associated production will have smaller 
rates for $t \bar t A$ compared to $t \bar t H$~(see for example 
\cite{BhupalDev:2007is,Boudjema:2015nda}). Hence production of the (pseudo) scalar in gluon 
fusion with associated jets will be a better channel to probe for such an invisibly decaying 
$A$. Clearly, more studies are required.

\section{Complementarity with dark matter searches}
\label{sec:DM}

\begin{figure}[h!]
\begin{center}
\begin{tabular}{c}
\hspace{-12mm}\includegraphics[scale=0.95]{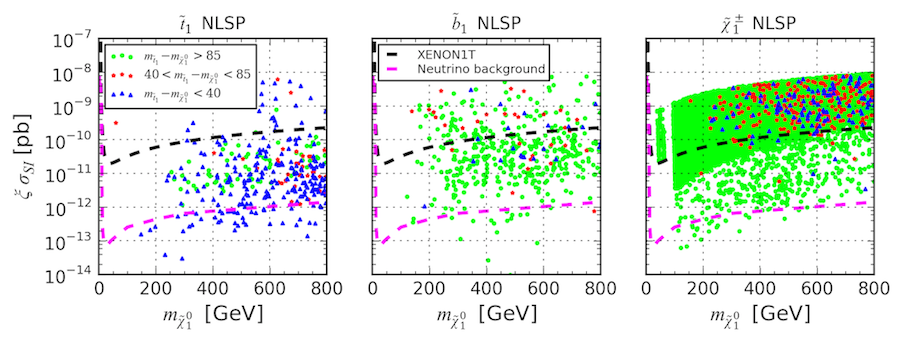}
\end{tabular}
\caption[]{ Spin-Independent $\neutl$-nucleon cross section vs the $\neutl$ mass for $\stopl$ (left), 
$\sbottom1$ (center)and $\charginol$ (right) NLSP. For comparison,  the expected limits from 
XENON1T and the neutrino coherent scattering are superimposed. 
\label{fig:res-2}}
\end{center}
\end{figure}

The spin-independent (SI) neutralino-proton cross section is displayed in Fig.~\ref{fig:res-2} 
for the cases of the $\stopl,\sbottom1$ and $\charginol$ NLSPs. The points are color 
coded according to the stop-LSP mass difference. Although this quantity is not directly relevant 
for direct detection it is useful to highlight the complementarity with collider searches. Here 
the cross section is rescaled according to Eq.~\ref{relic-limits} to take into account scenarios 
where the neutralino is only a part of the DM.

The results can be understood knowing that the predictions for the SI cross section are 
basically governed by the nature of the neutralino, pure states leading to small cross sections 
and mixed Higgsino/gaugino to the largest. A large fraction of the points will be probed by 
Xenon1T, in particular those with a chargino NLSP - which constitute the largest sample - since 
they are typically associated with a dominantly, yet mixed, Higgsino(wino) LSP. The right panel 
of Fig.~\ref{fig:res-2} shows that Xenon1T has the potential to cover the vast majority of 
points where 40 GeV $< \mstl-\mlsp < 85$~GeV. Those are the points that are 
not well constrained by current LHC bounds because the stop decays mostly into 
$b\charginol$ and the chargino in turn decays via a virtual W.  Some of the points with 
$\mstl-\mlsp < 40$~GeV can also be probed by Xenon1T provided the 
chargino is the NLSP. Furthermore the scenarios with a LSP with a mass around 45 or 60~GeV 
(corresponding to the so-called bino branches mentioned in previous sections) should be entirely 
probed with Xenon-1T. A few points with chargino NLSP lie below the coherent neutrino scattering 
background, these are typically associated with a pure Higgsino or wino LSP.

Less promising for direct detection are scenarios with a squark NLSP. Figure~\ref{fig:res-2}, 
left and center panels, shows that only a small fraction of the points will be probed by the 
future Xenon1T detector, a few points even lying below the irreducible coherent neutrino 
scattering background. In particular some of the points where 
$\mstl-\mlsp < 40$~GeV, in the left panel of Fig.~\ref{fig:res-2}, can 
lead to a very small cross section. The reason is that for such mass splitting the value of the 
relic density is governed by the coannihilation channels with stops, hence a dominantly bino LSP 
is allowed. Its weak coupling to the Higgs lead to a suppressed SI cross section.  Thus these 
points that are hard to probe at the LHC can also evade direct DM searches. 
Note that going beyond the standard cosmological scenario by assuming that DM can be regenerated ,
for example from decay of moduli fields re-injecting neutralinos after the 
freeze-out~\cite{Moroi:1999zb,Hall:2009bx,Arcadi:2011ev}, such that $\xi=1$ would lead to much stronger 
constraint~\cite{Williams:2012pz}.
Many of the scenarios would already be constrained by LUX and nearly all of the ones with chargino 
NLSP could be probed at Xenon1T.

\begin{figure}[h!]
\begin{center}
\begin{tabular}{c}
\hspace{-12mm}\includegraphics[scale=0.78]{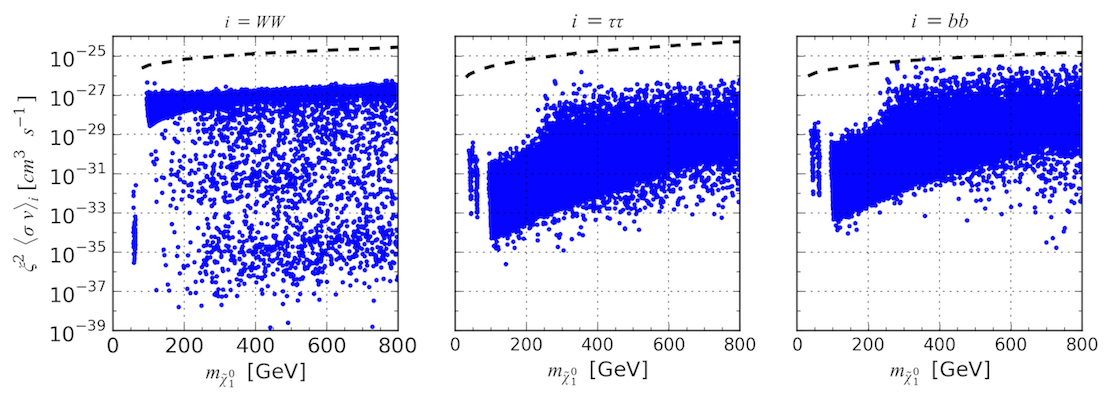}
\end{tabular}
\caption[]{Indirect detection cross section as a function of LSP mass in the $WW$, $\tau \tau$ 
and $bb$ final states . For comparison the current FermiLAT~\cite{Ackermann:2015zua} limit is 
superimposed.
\label{fig:indirect-xsec}}
\end{center}
\end{figure}

We have also computed the indirect detection cross section for LSP annihilation into 
$\tau^+\tau^-$, $b\bar{b}$ and $W^+W^-$ and compared this with the exclusion obtained by 
FermiLAT from observations of the photon flux from dwarf spheroidal galaxies (dSphs) of the 
Milky Way~\cite{Ackermann:2015zua}. Fig.~\ref{fig:indirect-xsec} shows the results for 
different channels after rescaling by $\xi^2$. Since a large fraction of the points have $\xi < 
1$, the rescaled cross section is often strongly suppressed, hence only a handful of points in 
the $b\bar{b}$ channel are excluded by such searches - basically those where the cross section 
is enhanced by annihilation through a heavy Higgs. Again assuming $\xi=1$ would lead to a 
completely different picture, with a strong increase in the predictions of the cross sections. 
In particular $\sigma v$ for DM annihilation in the WW channel exceeds the FermiLAT limit for 
most of the scenarios with a LSP mass below 300GeV, in agreement with the results 
in~\cite{Williams:2012pz,Cohen:2013ama}. It was also shown that PAMELA limits from antiprotons can 
constrain such scenarios~\cite{Belanger:2012ta} and that the wino can be also effectively probed 
by FermiLAT searches for gamma-ray lines from neutralino annihilation into photon pairs (or 
$\gamma Z$)~\cite{Fan:2013faa}.
 
\section{Conclusions}

The discovery of a SM-like Higgs boson has strong implications for SUSY since a Higgs mass of 126 GeV requires large radiative corrections from the stop sector. This requires heavy stops and/or large mixing which is in conflict with the naturalness arguments. Stops therefore play a central role in SUSY and are a key ingredient in testing the naturalness of the MSSM. The LHC has performed many dedicated searches for stops. 
However, their production cross sections are small compared to the first two generation squarks, which leads to degraded LHC run1 limits on their masses. Hence, it is interesting to investigate to which extent light stops are still allowed and demonstrate various possible future probes of the resultant MSSM scenario at the LHC via not just the searches for the stops but also for other sparticles such as sbottoms, electroweakinos and even the heavy Higgs.

We first determined the regions of the pMSSM with ten free parameters compatible with a light  stop (specifically with a mass below 1.5~TeV) after taking into account current constraints from the Higgs mass, Higgs signal strengths, flavour physics, the upper limit on the neutralino relic density as well as DM direct detection. Flavour constraints are very restrictive since they are in tension with the Higgs mass. For example, constraints from $B_d \to X_s \gamma$, can become stricter with large $A_t$, which is what the Higgs mass requires. The PLANCK upper limit on the relic density and the direct search limits from LUX are also in tension, combined they favour either TeV scale LSP or a almost pure Higgsino or wino LSP. The latter implies a supersymmetric spectra with a chargino NLSP and small mass differences between the chargino and the neutralino LSP. Another possibility is a squark NLSP (in particular a stop) since coannihilation can be used to obtain the measured value for the relic density. Both cases entail that LHC searches for SUSY are difficult. 

We have then used \smodels for all the MSSM points allowed by all the previously mentioned constraints to find implications of the limits obtained by the LHC collaborations on supersymmetric particles with the simplified model framework. We have found  that stops below 500 GeV can be consistent with all the LHC searches (including direct  stop and sbottom production). The reason is either a compressed spectra or reduced branching ratios into the channels excluded with simplified models. This is our first main result.

Our results generally agree with those of~\cite{deVries:2015hva} which also thoroughly investigates the impact of LHC SUSY searches on the pMSSM parameter space and found that light stops were allowed. Our approach however differs in two ways. First in our choice of the set of free parameters. Most importantly, in contrast to us \cite{deVries:2015hva} allows light sleptons and gluinos. Second, we impose only an upper bound on the DM relic density leading to a large fraction of points with large higgsino or wino components.

Another important aspect of our analysis is the identification of final states which would, in principle, be capable of probing the points surviving after the application of LHC, flavor, dark matter and Higgs constraints. This  analysis can help determine how best to extend the search reach at the LHC, particularly for the low masses still allowed after LHC constraints. One of the most important missing topology corresponds to asymmetric stop decays - for example one of the pair produced stop decaying to $t\neutl$ and the other into $b\charginol$ with the chargino being invisible because it decays into the LSP and soft jets. Other possible  signatures which might improve the reach  involve the decay of squarks into a heavier neutralino or chargino which then decay to a LSP via virtual W/Z.

An extension of the reach in the region of small mass difference of the stop with the LSP entails using the jet and MET signature from a stop decaying into charm neutralino or via 4-body decays as done by the LHC collaborations. Although we have not used these channels as they played a minor role in our analysis, they should lead to strong constraints in the future. We also investigated whether the region near the kinematic boundary could be probed by considering associated production of stops with a Higgs. In principle tagging the Higgs allows to handle the background very well.  We find that with the Higgs mass constraint, the stop composition is such that the $\stopl\stopl h$ coupling is small. As a result  this cross section is small, even for small masses of the stops being considered, the suppression being dynamic and not so much due to kinematics. Hence the stop search in this channel  is challenging and needs more detailed studies. The associated stop stop jet process is  perhaps  a better option, for the region near the kinematic boundary, though 
potentially more complicated to analyze. Thus the subject of associated production of a stop pair with a jet or Higgs requires further detailed studies.

Alternative probes of light stop scenarios involve search for the heavier stop through its $\stoph \to \stopl h/Z$ decay. Indeed the branching ratios for these channels in particular, for the decay $\stoph \to \stopl Z$, are expected to be large because of the observed large Higgs mass. 

Interestingly, for the allowed points in our light stop scenario the heavy Higgs phenomenology is found to be very interesting. Searches of the heavy Higgses  provide additional channels to probe the model, either though conventional signatures in the $\tau\tau$ decay channel or taking advantage of decays into electroweakinos, including  invisible decays. Our MSSM scenarios, with no relationships between different gaugino masses, in fact allows for considerable values for these branching ratios and yet satisfy the LHC8 constraint on electroweakinos.

Finally we highlight the complementarity with DM searches, nearly  all points with the stop-NLSP mass difference below the W mass will be tested at Xenon1T provided the NLSP is a chargino, this means that one region that is hard to cover at the LHC via squark and electroweakino searches will nicely be probed by ton scale detectors. Unfortunately, it is much harder to cover the region with a stop NLSP with direct DM searches.

In this analysis we have explicitly rejected the long-lived particles, however in our initial sample a large fraction of the points involved long-lived charged particles, in particular charginos. Existing searches for long-lived particles constrain severely the dominantly wino charginos in certain mass ranges~\cite{belanger:1505}. Clearly improved analyses could provide a handle to probe this region of the parameter space left unexplored in this work.

Thus, we have shown that the light stops, being actively hunted at the LHC, can be probed by more than one means. While the LHC Run2 will bring interesting results and hopefully a BSM signal, possible ways of constraining the light stop scenario via indirect constraints should also be considered.
\section*{Acknowledgements}
DG acknowledges support from the European Research Council under the European Union's Seventh 
Framework Programme (FP/2007-2013) / ERC Grant Agreement no. 279972. DG, RMG and GB wish to thank 
the organizers of the Workshop on High Energy Physics Phenomenology, WHEPP  at Puri, India. 
DG would also like to thank the hospitality of LAPTh where part of this work was done.
RMG wishes to acknowledge support from the Department of Science and Technology, India under Grant 
No. SR/S2/JCB-64/2007 under the J.C. Bose  Fellowship scheme as well as support from the French 
ANR Project ``DMAstro-LHC'',  ANR-12-BS05-0006, for a visit to LAPTh. SuK is supported by the 
``New Frontiers" program of the Austrian Academy of Sciences. We thank Jonathan Da Silva for providing 
us with the neutrino coherent scattering and Fermi limits. We thank Sabine Kraml, Andre Lessa, Ursula 
Laa, Veronika Magerl, Wolfgang Magerl, Michael Traub and Wolfgang Waltenberger for useful discussions.
\providecommand{\href}[2]{#2}\begingroup\raggedright\endgroup

\end{document}